\documentclass{article}
\RequirePackage{amsmath,amsfonts,amssymb}
\usepackage{graphicx}
\usepackage[longnamesfirst,round]{natbib}
\usepackage{makecell}
\usepackage{multirow,rotating}
\usepackage[h]{esvect}
\usepackage{hyperref}

\usepackage[letterpaper, portrait, margin=1in]{geometry}

\newcommand{\e}{\mathbb{E}}
\newcommand{\var}{\mathbb{V}}

\newcommand{\bs}{\boldsymbol}
\newcommand{\mbf}{\mathbf}

\newcommand{\bZ}{\bs{Z}}
\newcommand{\bU}{\bs{U}}

\newcommand{\bzero}{\bs{0}}

\newcommand{\bpi}{\bs{\pi}}

\newcommand{\momega}{\mbf{\Omega}}

\newcommand{\reals}{\mathbb{R}}

\newcommand{\bull}{\text{\raisebox{1pt}{\scalebox{.6}{$\bullet$}}}}

\begin{document}

\title{Toward improved inference for Krippendorff's Alpha agreement coefficient}

\author{John Hughes\\
Lehigh University\\
Bethlehem, PA, USA 18015}

\maketitle

\begin{abstract}
In this article I recommend a better point estimator for Krippendorff's Alpha agreement coefficient, and develop a jackknife variance estimator that leads to much better interval estimation than does the customary bootstrap procedure or an alternative bootstrap procedure. Having developed the new methodology, I analyze nominal data previously analyzed by Krippendorff, and two experimentally observed datasets: (1) ordinal data from an imaging study of congenital diaphragmatic hernia, and (2) United States Environmental Protection Agency air pollution data for the Philadelphia, Pennsylvania area. The latter two applications are novel. The proposed methodology is now supported in version 2.0 of my open source R package, \texttt{krippendorffsalpha}, which supports common and user-defined distance functions, and can accommodate any number of units, any number of coders, and missingness. Interval computation can be parallelized.
\end{abstract}

\section{Introduction}

Krippendorff's $\alpha$ \citep{hayes2007answering} is a well-known methodology for statistically assessing agreement. Although $\alpha$ is non-parametric, the customary $\alpha$ estimator is motivated by an estimator of the intraclass correlation coefficient in the one-way mixed-effects analysis of variance (ANOVA) model \citep{ravishanker2021first}, a much studied fully parametric model. In this article I leverage $\alpha$'s connection with the one-way mixed-effects ANOVA model to explore the customary approach to inference for Krippendorff's $\alpha$, finding that both the point estimator and interval estimation have substantial drawbacks for smaller, yet realistic, sample sizes. Then I consider a better point estimator; and propose a jackknife variance estimator \citep{hinkley1977jackknife} that yields interval estimates having very nearly their desired coverage rates, even in unfavorable conditions. I evaluate the various procedures not only formally but also by way of extensive and realistic simulation studies. Finally, I analyze data previously analyzed by Krippendorff and others, along with two experimentally observed datasets: (1) radiologist-assigned grades in an imaging study of congenital diaphragmatic hernia (CDH), and (2) United States Environmental Protection Agency (EPA) PM$_{2.5}$ data from seven geographically dispersed air sensors in or near Philadelphia, Pennsylvania.

\section{Measuring agreement}

An inter-coder agreement coefficient---which takes a value in the unit interval, with 0 indicating no agreement and 1 indicating perfect agreement---is a statistical measure of the extent to which two or more coders agree regarding the same units of analysis. The agreement problem has a long history and is important in many fields of inquiry, and numerous agreement statistics have been proposed.

The earliest agreement coefficients were $S$ \citep{sagree}, $\pi$ \citep{scottpi}, and $\kappa$ \citep{cohenkappa}. \citet{sagree} proposed the $S$ score as a measure of the extent to which two methods of communication provide identical information. \citet{scottpi} proposed the $\pi$ coefficient for measuring agreement between two coders. \citet{cohenkappa} criticized $\pi$ and proposed the $\kappa$ coefficient as an alternative to $\pi$---although \citet{smeeton} noted that Francis Galton mentioned a $\kappa$-like statistic in his 1892 book, {\em Finger Prints}. \citet{fleiss1971} proposed multi-$\kappa$, a generalization of Scott's $\pi$ for measuring agreement among more than two coders. \citet{conger1980} and \citet{davies1982} likewise generalized $\kappa$ to the multi-coder setting. Other generalizations of $\kappa$, e.g., weighted $\kappa$ \citep{weightedkappa}, have also been proposed. The $\kappa$ coefficient and its generalizations can fairly be said to dominate the field and are still widely used despite their well-known shortcomings \citep{feinstein1990high,cicchetti1990high}.

Other oft-used measures of agreement are Gwet's $AC_1$ and $AC_2$ \citep{gwet2008computing} and Krippendorff's $\alpha$ \citep{hayes2007answering}, the latter of which is the subject of this article. An even newer agreement methodology is Sklar's $\omega$ \citep{sklarsomega}, a parametric Gaussian copula-based framework. For more comprehensive reviews of the literature on agreement, I refer the interested reader to the article by \citet{banerjee1999beyond}, the article by \citet{artstein2008inter}, and the book by \citet{gwet}.

\section{A motivating example}

To fix ideas, let us consider an example dataset that was previously analyzed by \citet{krippendorff2013}. The dataset, which comprises 41 nominal codes assigned to a dozen units of analysis by four coders, is shown below. The dots represent missing values.

\begin{figure*}[h]
   \begin{center}
   \begin{tabular}{ccccc}
   & $c_1$ & $c_2$ & $c_3$ & $c_4$ \\
  $u_1$ &   1  &  1  & \bull  &  1 \\
 $u_2$ &    2  &  2  &  3  &  2 \\
 $u_3$ &   3  &  3 &   3 &   3 \\
 $u_4$ &    3  &  3 &   3  &  3 \\
$u_5$ &    2  &  2 &   2 &   2 \\
 $u_6$ &   1  &  2 &   3 &   4 \\
 $u_7$ &    4  &  4  &  4  &  4 \\
$u_8$ &   1  &  1  &  2  &  1 \\
 $u_9$ &   2  &  2  &  2  &  2 \\
$u_{10}$ &   \bull &    5  &  5  &  5 \\
$u_{11}$ &   \bull &   \bull&    1  &  1 \\
$u_{12}$ &   \bull  &  3  & \bull  &  \bull
   \end{tabular}
   \end{center}
   \caption{Nominal scores previously analyzed by Krippendorff, for twelve units and four coders. The dots represent missing values.}
   \label{fig:nominal}
\end{figure*}

Because this dataset is small and the codes are nominal, it is easy to hypothesize by inspection that agreement is high. Indeed, eight of the units exhibit perfect agreement, and two of the remaining units exhibit near-perfect agreement. The only unit about which the coders evidently disagreed is unit 6. And of course the final unit carries no information regarding agreement. These facts taken together suggest that an estimated agreement coefficient for these data should not be too far from 1, unless the estimator in question is strongly influenced by the disagreement over unit 6.

Before analyzing these data I should mention that I will interpret results according to the agreement scale given in Table~\ref{interpret} \citep{landiskoch}. Although this scale is well-established, agreement scales remain a subject of debate \citep{taber2018}, and so the following scale---indeed, any agreement scale---should be applied circumspectly.

\begin{table}[h]
\caption{Guidelines for interpreting values of an agreement coefficient.} 
\label{interpret}
\begin{center}
\begin{tabular}{cl}
Range of Agreement & Interpretation\\\cline{1-2}
$\phantom{0.2<\;}\alpha\leq 0.2$ & Slight Agreement\\
$0.2<\alpha\leq 0.4$ & Fair Agreement\\
$0.4<\alpha\leq 0.6$ & Moderate Agreement\\
$0.6<\alpha\leq 0.8$ & Substantial Agreement\\
$\phantom{0.2<\;}\alpha>0.8$ & Near-Perfect Agreement
\end{tabular}
\end{center}
\end{table}

Applying the customary Krippendorff's $\alpha$ methodology to these data, with the discrete metric $d^2(x,y)=1\{x\neq y\}$ as the distance function, yields point estimate $\hat{\alpha}=0.743$ and 95\% confidence interval $(0.459,1.000)$. This estimate of $\alpha$ indicates substantial agreement, and the interval suggests that these data are consistent with agreement ranging from fair to perfect. If one repeats the analysis having removed unit 6, the point estimate changes to $0.857$, and the interval becomes $(0.679,1.000)$. Thus we see that unit 6 was (perhaps unduly) influential since the new results indicate near-perfect agreement (point estimate) and at least substantial agreement (interval estimate).

I will return to these data in Section~\ref{realdata}, where I will apply my proposed methodology and compare those results to these.

\section{The customary Krippendorff's $\alpha$ methodology}

\citet{kripppkg} showed that Krippendorff's $\alpha$ finds its origin in the well-known one-way mixed-effects ANOVA model. In this section I will review \citeauthor{kripppkg}' demonstration, and then elaborate on it for the purposes of this article.

\subsection{Krippendorff's $\alpha$ and the one-way mixed-effects ANOVA model}

The one-way mixed-effects ANOVA model is given by
\[
Y_{ij}=\mu+\tau_i+\varepsilon_{ij},\;\;\;\;(i=1,2,\dots, a)\;(j=1,2,\dots,n_i)
\]
where
\begin{itemize}
\item $Y_{ij}$ is the $j$th score (of $n_i$ scores) for the $i$th unit (of $a$ units of analysis);
\item $\mu\in\reals$ is the population mean score;
\item $\tau_i\stackrel{\text{ind}}{\sim}\textsc{Normal}(0,\sigma_\tau^2)$ are random unit effects such that $\sigma_\tau^2\geq 0$;
\item $\varepsilon_{ij}\stackrel{\text{ind}}{\sim}\textsc{Normal}(0,\sigma_\varepsilon^2)$ are errors such that $\sigma_\epsilon^2>0$; and
\item the unit effects are independent of the errors.
\end{itemize}
Since the scores for the $i$th unit share the unit effect $\tau_i$, said scores are dependent. Specifically, for $j\neq j'$, $\text{cov}(Y_{ij},Y_{ij'})=\sigma_\tau^2$, and
\begin{align}
\label{icc}
\alpha&=\text{cor}(Y_{ij},Y_{ij'})=\frac{\sigma_\tau^2}{\sigma_\tau^2+\sigma_\varepsilon^2}.
\end{align}
This correlation among the scores for a given unit is usually called the intraclass correlation coefficient (ICC). I denote the ICC as `$\alpha$' precisely because the ICC is the population parameter for Krippendorff's $\alpha$ when the data conform to the one-way mixed-effects ANOVA model. To reveal this connection it suffices to show that Krippendorff's estimator, which I denote as $\hat{\alpha}$, is an estimator of $\alpha$.

First, note that $\alpha$ can be written as
\[
\alpha = 1 - \frac{\sigma_\varepsilon^2}{\sigma_\tau^2+\sigma_\varepsilon^2}.
\]
This suggests the estimator
\[
\hat{\alpha}=1-\frac{\widehat{\sigma_\varepsilon^2}}{\widehat{\sigma_\tau^2+\sigma_\varepsilon^2}},
\]
which we can completely specify by identifying estimators $\widehat{\sigma_\varepsilon^2}$ and $\widehat{\sigma_\tau^2+\sigma_\varepsilon^2}$. For the one-way mixed-effects ANOVA model, the customary estimator of the error variance $\sigma_\varepsilon^2$ is the so called mean squared error ($MSE$):
\[
\widehat{\sigma_\varepsilon^2}=MSE=\frac{SSE}{N-a}=\frac{\sum_{i=1}^a\sum_{j=1}^{n_i}(Y_{ij}-\bar{Y}_{i\bull})^2}{N-a},
\]
where $SSE$ denotes the error sum of squares, $N=\sum_in_i$ is the total sample size, and $\bar{Y}_{i\bull}$ is the sample mean for the $i$th unit. $MSE$ is both the method of moments (MoM) estimator and the maximum likelihood estimator of the error variance for the balanced design (i.e., when $n_i=n$ for all $i$). For the unbalanced design, $MSE$ is once again the MoM estimator of $\sigma_\varepsilon^2$, but the maximum likelihood estimator of $\sigma_\varepsilon^2$ is not available in closed form. For both designs $MSE$ is unbiased for $\sigma_\varepsilon^2$.

Now, to estimate the total variance $\sigma_\tau^2+\sigma_\varepsilon^2$, Krippendorff uses
\[
\widehat{\sigma_\tau^2+\sigma_\varepsilon^2}=MST_c=\frac{SST_c}{N-1}=\frac{\sum_{i=1}^a\sum_{j=1}^{n_i}(Y_{ij}-\bar{Y}_{\bull\bull})^2}{N-1},
\]
where $SST_c$ denotes the corrected (for the population mean) total sum of squares and $\bar{Y}_{\bull\bull}$ denotes the mean for the entire sample. This estimator seems quite natural given that
\[
\e MST_c=\frac{N-\frac{\sum_in_i^2}{N}}{N-1}\sigma_\tau^2+\sigma_\varepsilon^2\approx\sigma_\tau^2+\sigma_\varepsilon^2,
\]
with equality only when $\sigma_\tau^2=0$ (or $n_i=1$ for all $i$, which makes no sense). In any case, we arrive at Krippendorff's point estimator:
\begin{align}
\label{akripp}
\hat{\alpha}&=1-\frac{MSE}{MST_c}.
\end{align}

This estimator is the customary estimator for Krippendorff's $\alpha$ when squared Euclidean distance $d^2(x, y)=(x-y)^2$ is employed as the measure of discrepancy. \citet{kripppkg} showed how this form of $\hat{\alpha}$ can give rise to the non-parametric form of Krippendorff's $\alpha$, which is incidentally a modified multi-response permutation procedure \citep{permutation}. The non-parametric form of $\alpha$ simply makes $d^2$ a parameter whose value is chosen by the practitioner based on the type of outcomes to be analyzed---e.g., the discrete metric $d^2(x,y)=1\{x\neq y\}$ for nominal observations, distance function $d^2(x,y)=\{(x-y)/(x+y)\}^2$ for ratio observations, etc.

It is important to note that $\alpha$ is an agreement coefficient for all types of outcomes and suitable distance functions $d^2$. However, $\alpha$ is not a well-defined population parameter for every sensible choice of $d^2$. For example, when the observations are categorical and the discrete metric is used, $\hat{\alpha}$ is surely an estimator of agreement, but the population parameter that $\hat{\alpha}$ estimates cannot be described precisely. This reminds one of the $g$ factor \citep{warne2019spearman}, a construct that has been defined operationally as that which is measured by various cognitive tests.

\subsection{Bias of the customary point estimator}

\begin{figure}[ht]
   \begin{center}
   \begin{tabular}{ccc}
   \includegraphics[scale=.4]{long.jpg}\;\;\; $16\times 4$ & \includegraphics[scale=.4]{square.jpg}\;\;\; $8\times 8$ & \includegraphics[scale=.4]{short.jpg}\;\;\; $4\times 16$\\
   \includegraphics[scale=.23]{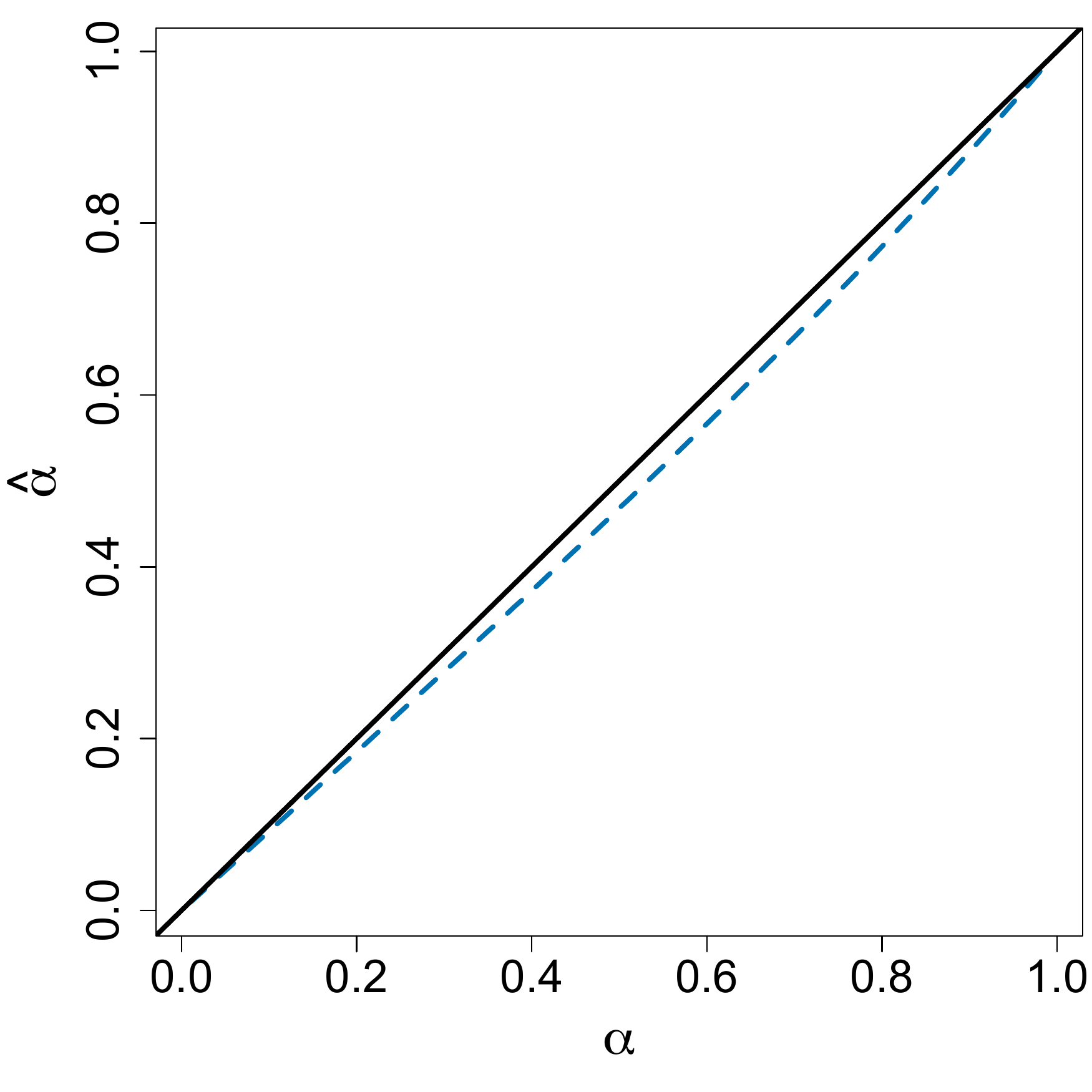} & \includegraphics[scale=.23]{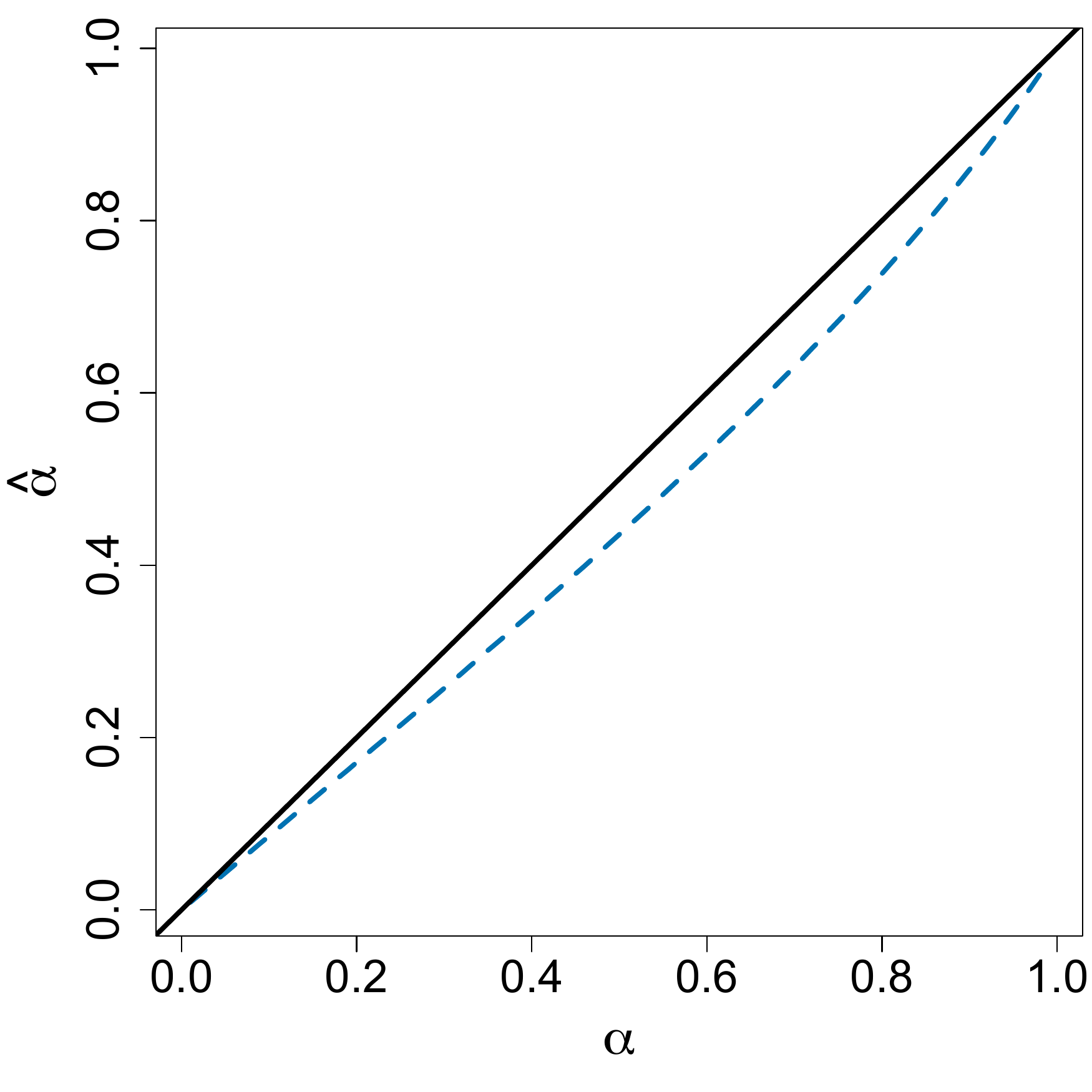} & \includegraphics[scale=.23]{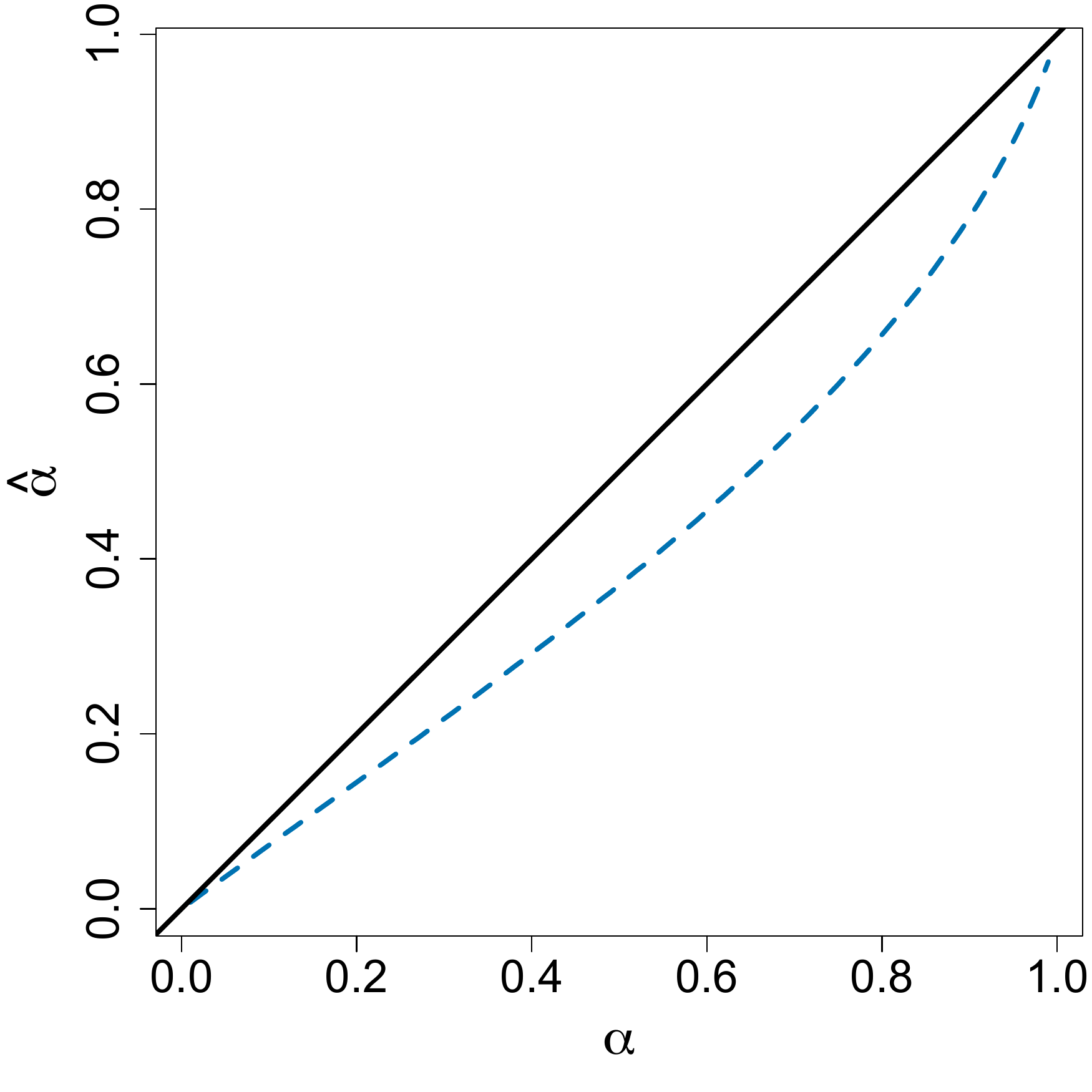} \\
   \includegraphics[scale=.23]{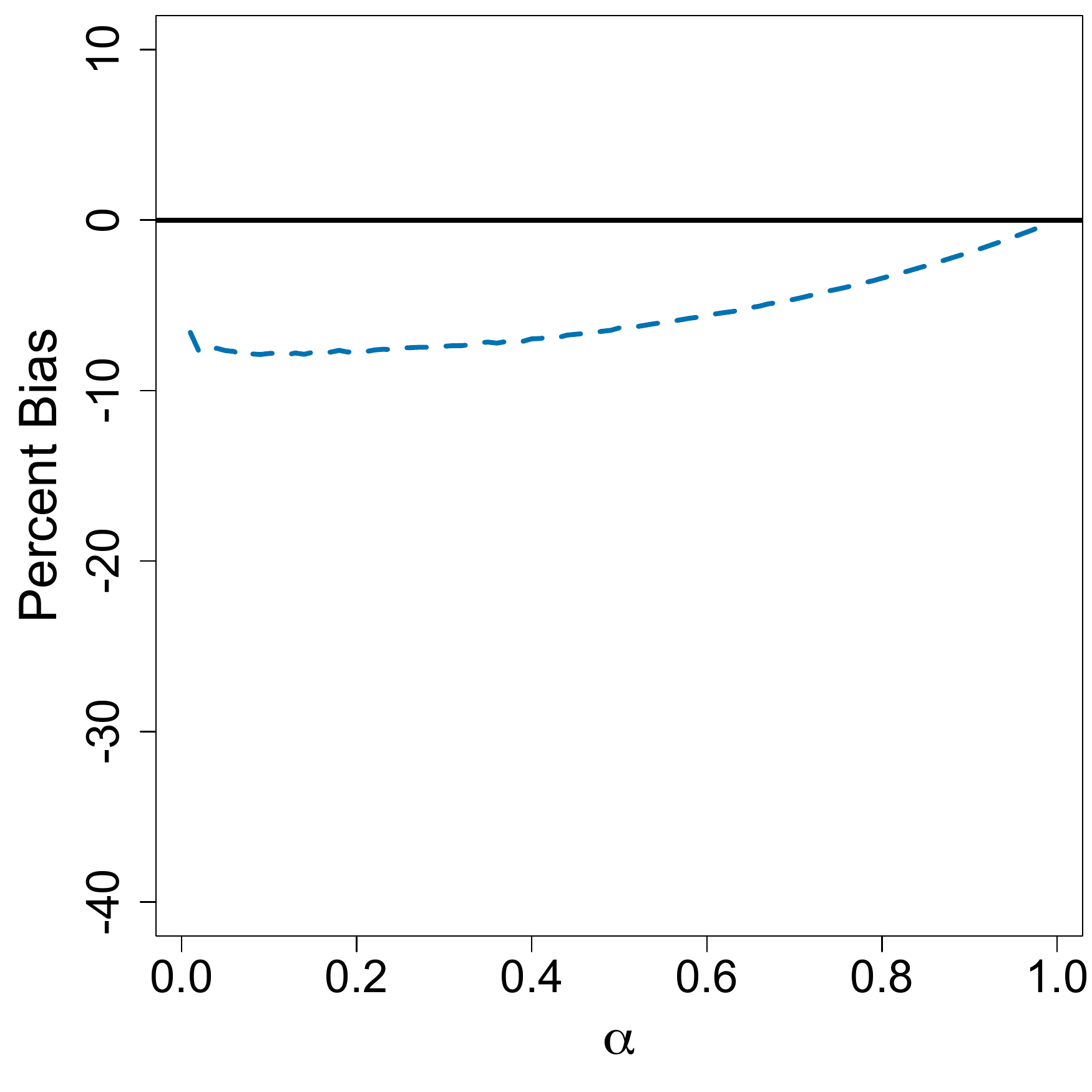} & \includegraphics[scale=.23]{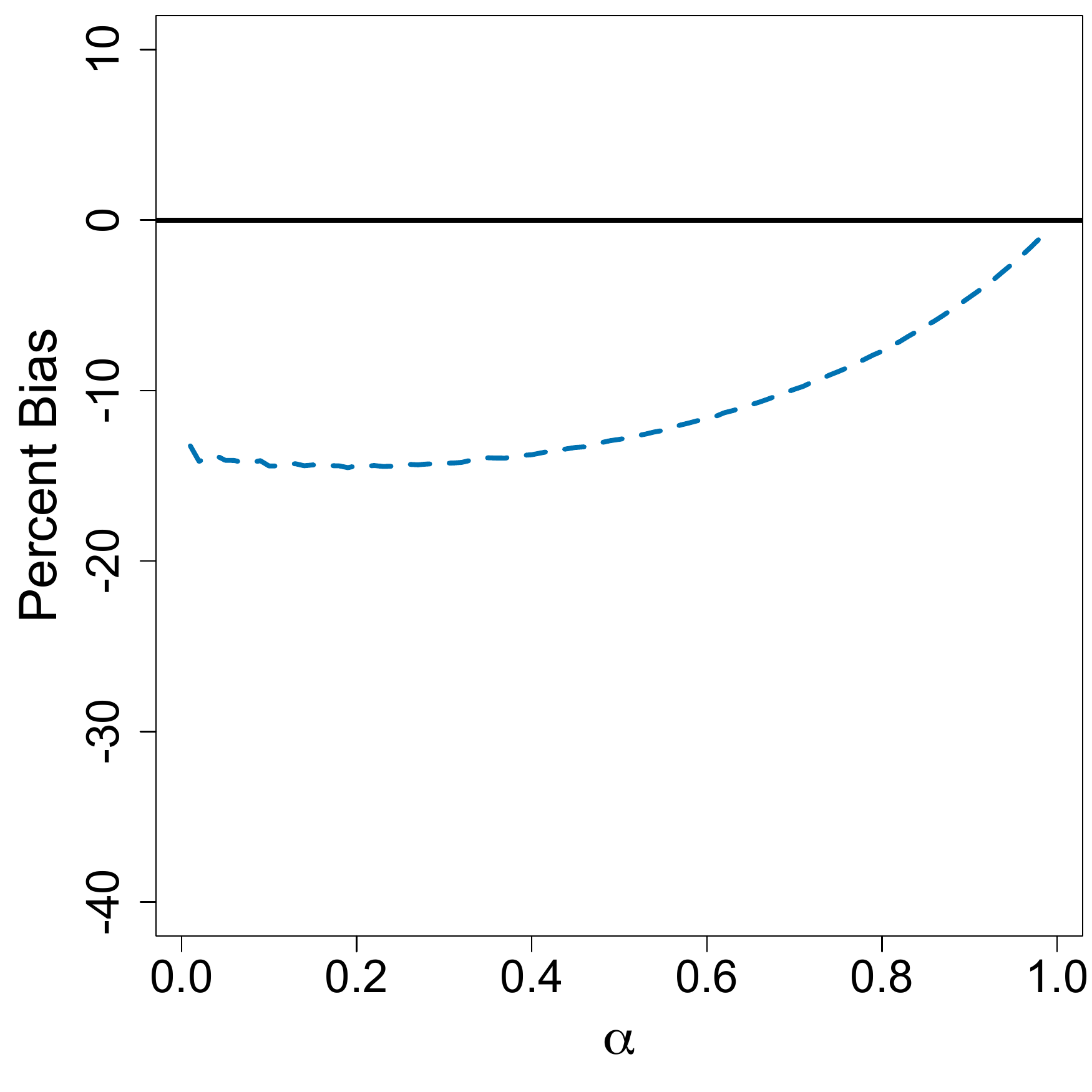} & \includegraphics[scale=.23]{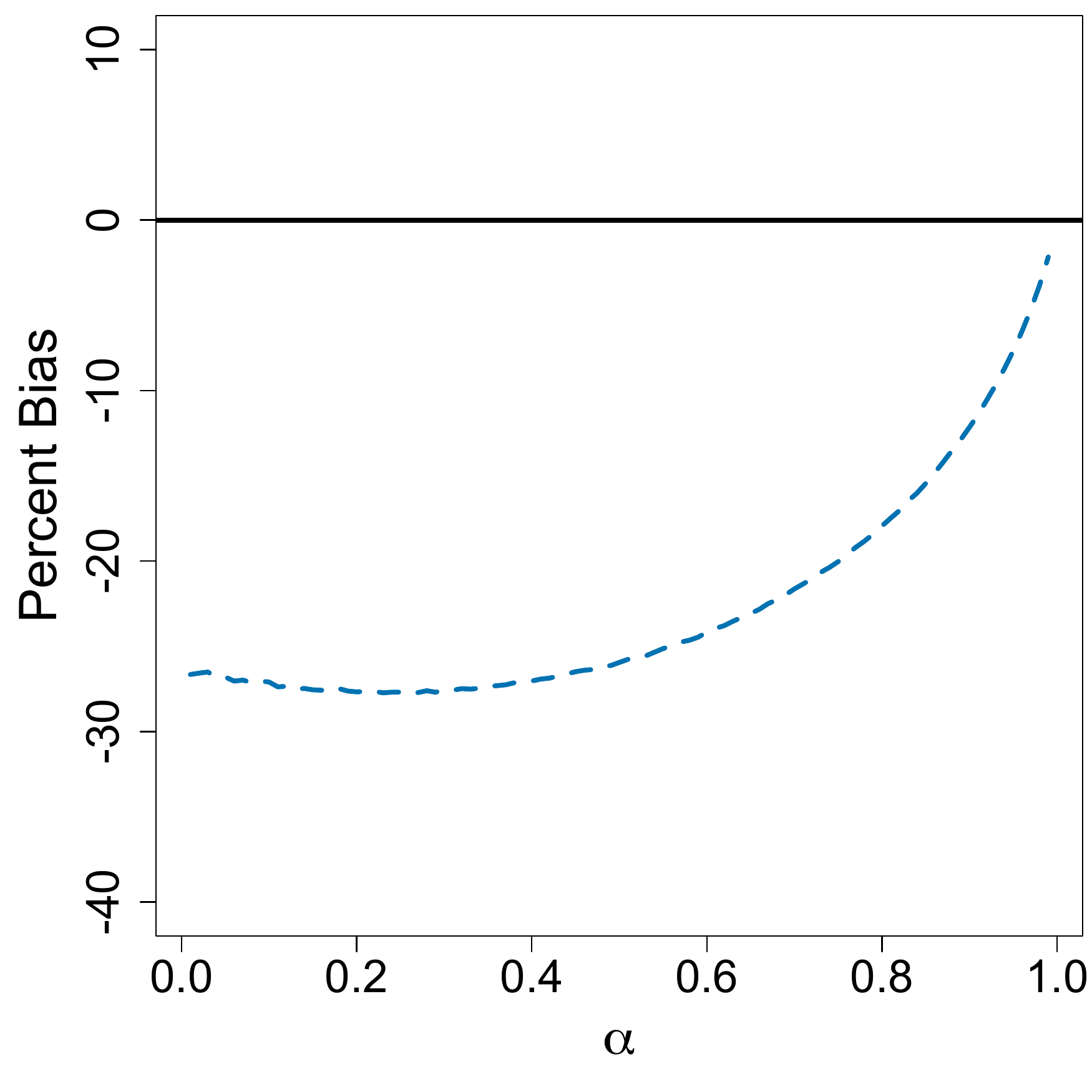}
   \end{tabular}
   \end{center}
   \caption{The raw bias (top row) and percent bias (bottom row) for the customary Krippendorff's $\alpha$ point estimator.}
   \label{customarybias}
\end{figure}

Note that $MST_c$ is biased downward, and the magnitude of the bias grows as the (average) number of coders increases (for fixed $N$). This implies that $\hat{\alpha}$, which already has a negative bias, becomes much more biased as the shape of the data matrix goes from tall to square to short (\includegraphics[scale=.3]{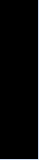} $\to$ \includegraphics[scale=.3]{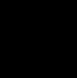} $\to$ \includegraphics[scale=.3]{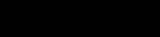}). This is shown in Figure~\ref{customarybias}, where the simulated outcomes were Gaussian and three balanced designs were used: (1) 16 units and 4 coders, (2) 8 units and 8 coders, and (3) 4 units and 16 coders.

We see that, as the aspect ratio of the data matrix increases from $1/4$ to 1 to 4, the percent bias increases dramatically. For the $16\times 4$ data matrix the maximum percent bias is nearly 10\%. The maximum percent bias then increases to approximately 15\% and 30\% for the square and short matrices, respectively. This unappealing behavior can be remedied (Section~\ref{better}).

\subsection{Coverage rates of the customary interval estimator}

The customary approach to interval estimation for Krippendorff's $\alpha$ employs a bootstrapping procedure \citep{alphaboot} that seems intuitive but struggles to perform well. The bootstrap sample is produced by computing
\[
\hat{\alpha}_k^*=1-\frac{MSE_k^*}{MST_c}
\]
for $k=1,\dots, b$, where $b$ is the desired bootstrap sample size; $MSE_k^*$ is $MSE$ computed for a data matrix that was created by resampling, with replacement, the rows of the observed data matrix; and $MST_c$ is the total mean square for the observed data matrix. Given the resulting bootstrap sample $\hat{\alpha}_1^*,\dots,\hat{\alpha}_b^*$, one estimates a confidence interval for $\alpha$ by computing the appropriate quantiles of the bootstrap sample \citep{efron1982jackknife}. That is, we compute a $(1-\delta)100\%$ interval as
\[
(L=\hat{\alpha}_{(\delta/2)}^*,\;U=\hat{\alpha}_{(1-\delta/2)}^*),
\]
where $\delta\in(0,1)$ is the desired significance level. Note that the percentile method is used here because the distribution of $\hat{\alpha}$ tends to be skewed, especially as $\alpha$ approaches 0 or 1.

The rationale for this method of interval estimation is two-fold. First, the resampling procedure leaves the rows intact to avoid breaking the dependence we aim to measure. Resampling the whole dataset would lead to much-inflated bootstrap estimates of $\sigma_\epsilon^2$, which would lead to a badly negatively biased bootstrap sample of $\hat{\alpha}_k^*$. Second, $MST_c$ is held fixed because the total sum of squares is invariant to permutation of the data.

It turns out that both of these arguments are incorrect. The second argument---that $MST_c$ should be held fixed---has the most deleterious effect on interval coverage rates. Of course it is true that $MST_c$ is invariant to permutation of the dataset, but that is irrelevant. What is relevant is that $MSE/MST_c$ is a ratio of estimators. Thus holding $MST_c$ fixed while using resampled data matrices to produce $MSE_1^*,\dots,MSE_b^*$ yields bootstrapped ratios $MSE_k^*/MST_c$ that are severely under-dispersed relative to the estimator $MSE/MST_c$. This is not surprising since failing to account for the variability of $MST_c$ as well as the variability of $MSE$ is analogous to erroneously assuming that a Student's $t$ distributed statistic is standard Gaussian. For smaller samples this, along with the bias of $\hat{\alpha}$, leads to confidence intervals that have abysmal coverage rates for nearly all values of $\alpha$.

\begin{figure}[ht]
   \begin{center}
   \begin{tabular}{ccc}
   \includegraphics[scale=.4]{long.jpg}\;\;\; $16\times 4$ & \includegraphics[scale=.4]{square.jpg}\;\;\; $8\times 8$ & \includegraphics[scale=.4]{short.jpg}\;\;\; $4\times 16$\\
   \includegraphics[scale=.23]{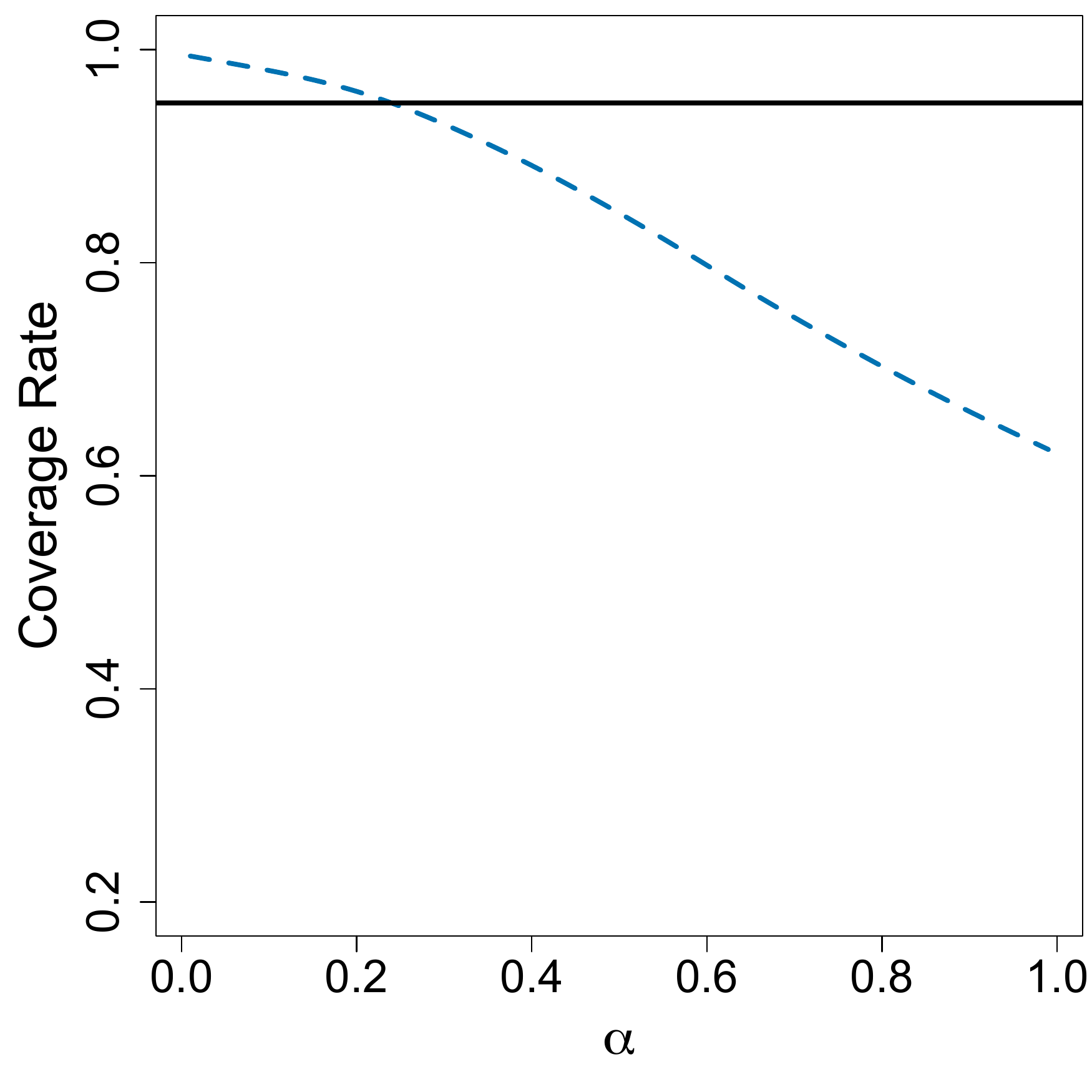} & \includegraphics[scale=.23]{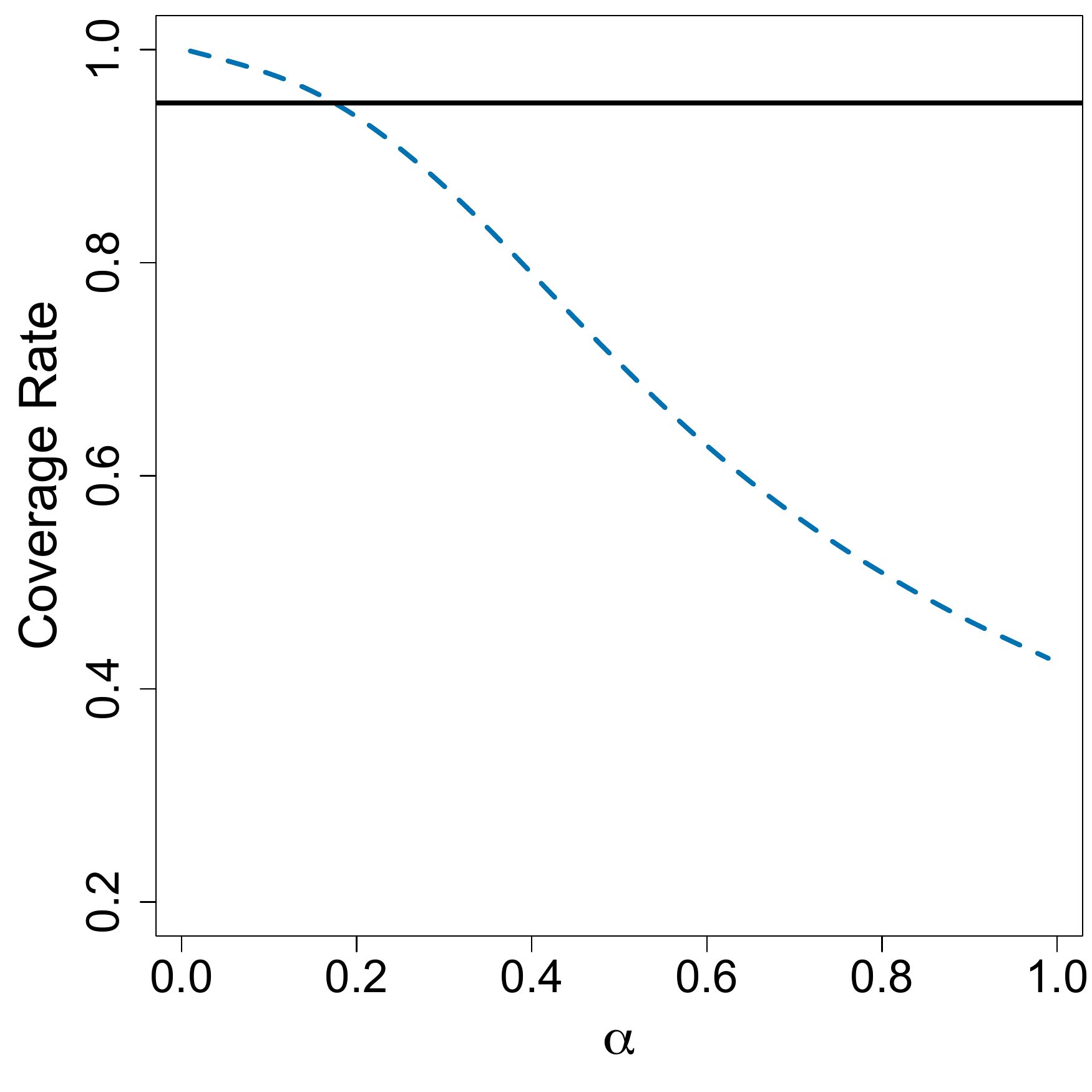} & \includegraphics[scale=.23]{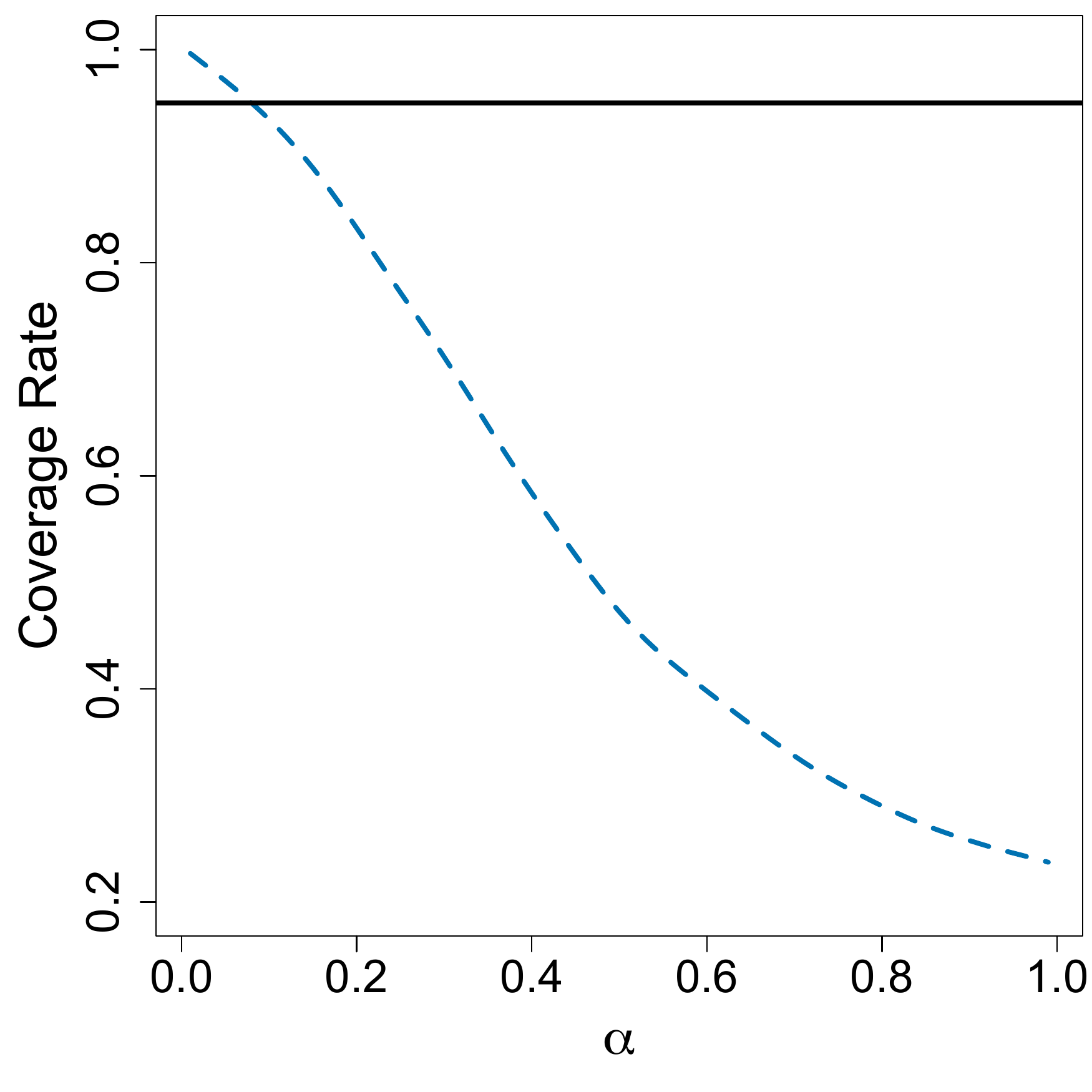} \\
   \end{tabular}
   \end{center}
   \caption{Empirical coverage rates of 95\% intervals computed using the customary Krippendorff's $\alpha$ point estimator and bootstrap procedure.}
   \label{customarycover}
\end{figure}

The plots in Figure~\ref{customarycover} show coverage rates for 95\% intervals that were computed using the methodology just described. Once again I include plots for tall, square, and short matrices such that $N=an$ remains constant at 64. We see that the customary interval estimation method performs well for small values of $\alpha$, but as $\alpha$ increases, performance degrades very rapidly and is generally unacceptable.

\subsection{Attempting to salvage bootstrap inference for $\alpha$}
\label{salvage}

As I implied in the preceding section, we should consider producing a bootstrap sample $\hat{\alpha}_1^*,\dots,\hat{\alpha}_b^*$ by resampling the rows of the data matrix with replacement and computing
\[
\hat{\alpha}_k^*=1-\frac{MSE_k^*}{(MST_c)_k^*}
\]
for each of $b$ resampled datasets. This would allow us to account for both sources of variation in $\hat{\alpha}=1-MSE/MST_c$.

While this more principled approach does very much improve coverage rates relative to the customary method, said rates are still much too low overall. This is attributable to the simple fact that, for data matrices having a small number of rows, resampling the rows with replacement tends to introduce so much redundancy, i.e., so many duplicate rows, that the apparent variation in the resampled dataset is much smaller than the variation of the original data. This once again leads to rather under-dispersed bootstrap samples, which leads to optimistic confidence intervals.

\begin{figure}[ht]
   \begin{center}
   \begin{tabular}{ccc}
   \includegraphics[scale=.4]{long.jpg}\;\;\; $16\times 4$ & \includegraphics[scale=.4]{square.jpg}\;\;\; $8\times 8$ & \includegraphics[scale=.4]{short.jpg}\;\;\; $4\times 16$\\
   \includegraphics[scale=.23]{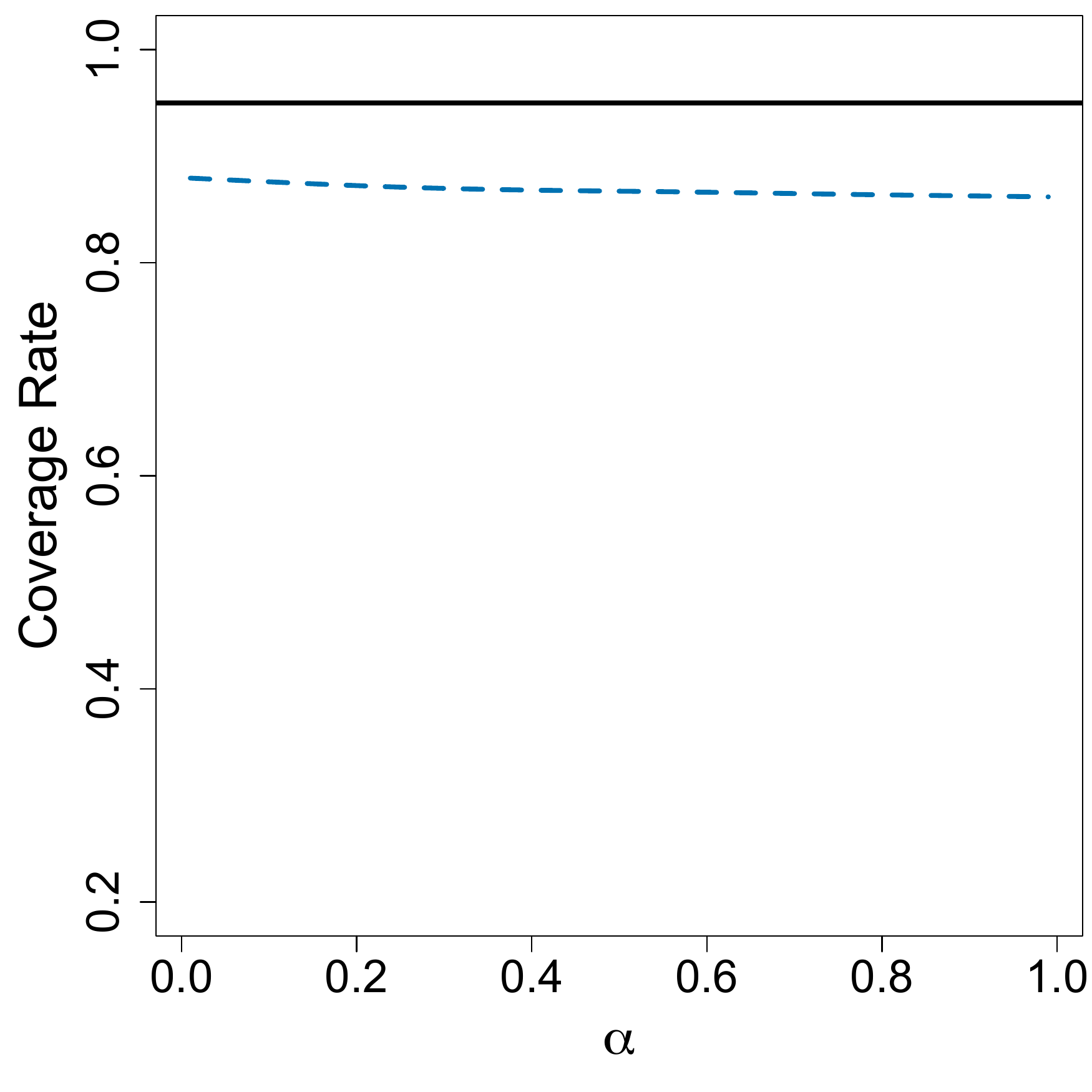} & \includegraphics[scale=.23]{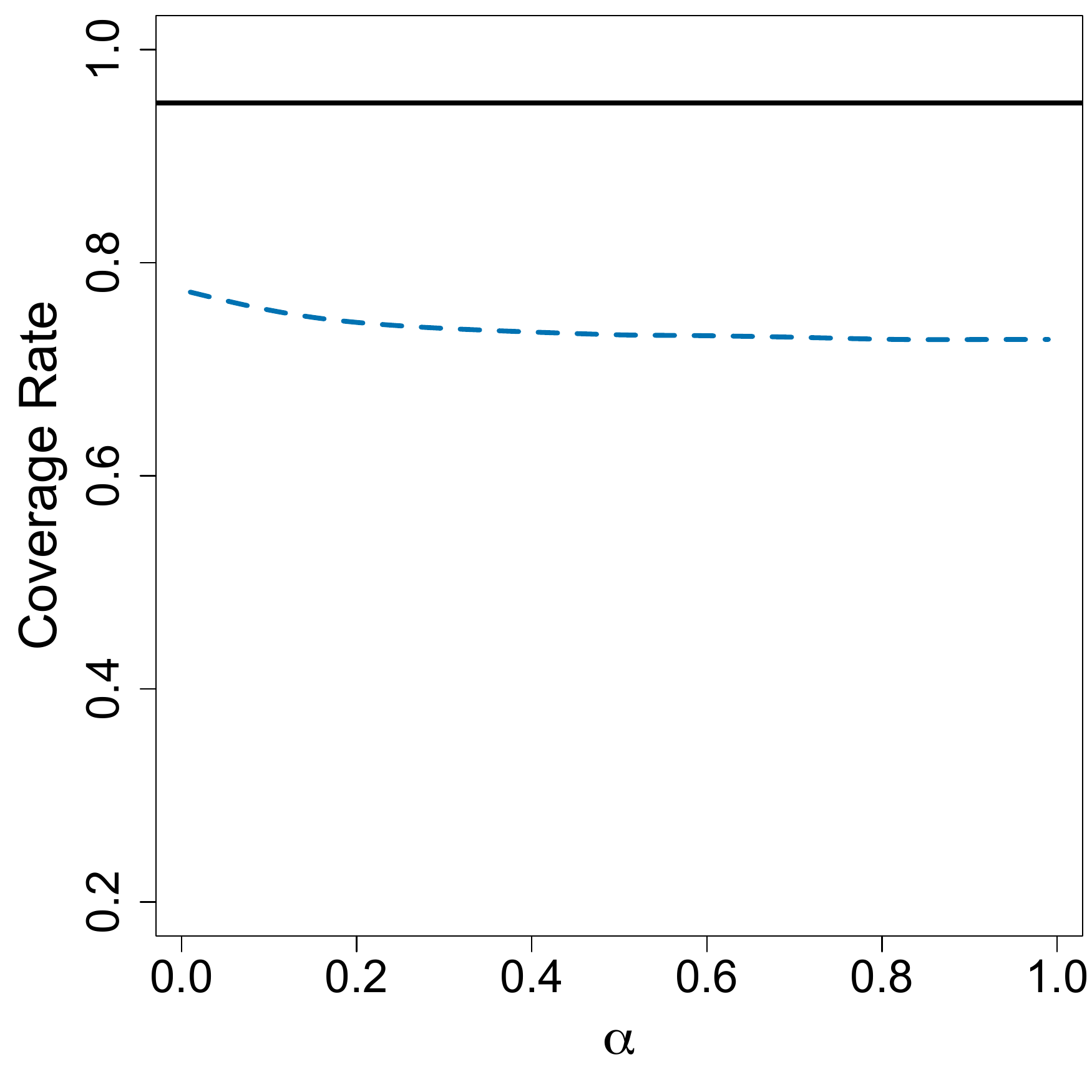} & \includegraphics[scale=.23]{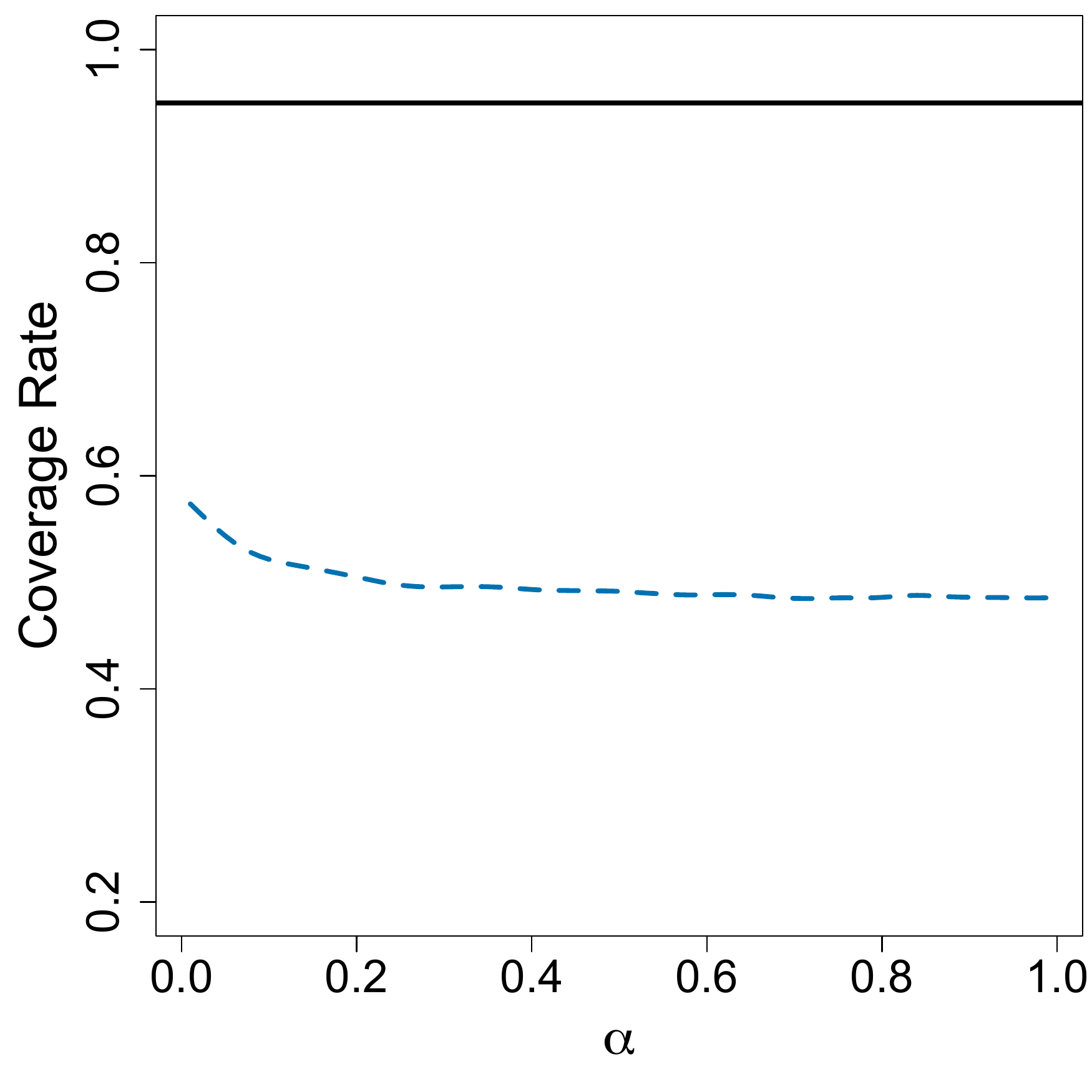} \\
   \end{tabular}
   \end{center}
   \caption{Empirical coverage rates of 95\% intervals computed using the customary Krippendorff's $\alpha$ point estimator and the improved bootstrap procedure.}
   \label{bettercover}
\end{figure}

The plots in Figure~\ref{bettercover} show coverage rates for 95\% intervals that were computed using the improved bootstrap procedure just described. This figure, too, includes plots for tall, square, and short matrices such that $N=an$ remains constant at 64. The improved bootstrap method performs much better overall than the customary method, yet coverage rates remain much too low.

\section{Alternative point estimators for $\alpha$}
\label{better}

The one-way mixed-effects ANOVA model, although apparently simple as a data-generating mechanism, has inspired a rich literature because the model is rather complicated as a data-analytic tool. In this section I describe five alternative estimators of $\alpha$ found in the literature. I compare the five estimators to one another and to $\hat{\alpha}$ by way of an extensive simulation study. Since the five estimators have been considered in some depth and detail elsewhere, I will keep my presentation brief.

Having chosen a suitable point estimator from among the six considered, I develop a corresponding interval estimation method that performs quite well, even for smaller and/or shorter data matrices. The proposed method also performs well computationally for smaller datasets, when the method is needed most.

\subsection{The maximum likelihood estimator}

The maximum likelihood estimator (MLE) of $\alpha$ is available in closed form for the balanced model but not for the unbalanced model. In the balanced case we have
\begin{align}
\label{mle}
\mathring{\alpha}&=\frac{(1-1/a)MSA-MSE}{(1-1/a)MSA + (n - 1)MSE},
\end{align}
where
\[
MSA=\frac{SSA}{a-1}=\frac{\sum_{i=1}^an_i(\bar{Y}_{i\bull}-\bar{Y}_{\bull\bull})^2}{a-1}
\]
is the ``treatment" mean square. The above estimator can be obtained by appealing to the invariance of maximum likelihood estimators and substituting the MLEs $\mathring{\sigma_\tau^2}=\{(1 - 1/a)MSA-MSE\}/n$ and $\mathring{\sigma_\varepsilon^2}=MSE$ for $\sigma_\tau^2$ and $\sigma_\varepsilon^2$ in (\ref{icc}). For Gaussian unit effects and errors, the estimator (\ref{mle}) is the MLE for $\alpha$ if and only if $\mathring{\sigma_\tau^2}$ is the MLE for $\sigma_\tau^2$, which requires that $\mathring{\sigma_\tau^2}$ be non-negative. When $\mathring{\alpha}$ is the MLE of $\alpha$, the estimator is not the ratio of unbiased estimators, nor is the MLE itself unbiased. But of course $\mathring{\alpha}$ possesses the attractive properties of MLEs, namely, consistency, statistical efficiency, and asymptotic normality.

\subsection{The analytical estimator}

The most commonly used estimator of $\alpha$, which is sometimes called the analytical estimator, is given by
\begin{align}
\label{common}
\tilde{\alpha} &= \frac{\tilde{\sigma_\tau^2}}{\tilde{\sigma_\tau^2}+\tilde{\sigma_\varepsilon^2}}=\frac{MSA-MSE}{MSA+(n-1)MSE},
\end{align}
where $MSE$ is unbiased for $\sigma_\varepsilon^2$ and $(MSA-MSE)/n$ is unbiased for $\sigma_\tau^2$. This is for a balanced model. The corresponding estimator for the unbalanced model is
\[
\frac{MSA-MSE}{MSA+(n^*-1)MSE},
\]
where
\[
n^*=\frac{N-\frac{1}{N}\sum_{i=1}^an_i^2}{a-1}
\]
is the average number of codes per unit of analysis. These are clearly plug-in MoM estimators since $\tilde{\sigma_\tau^2}$ and $\tilde{\sigma_\varepsilon^2}$ are MoM estimators and $\tilde{\alpha}$ is obtained by plugging these estimators into (\ref{icc}).

Although the form of $\tilde{\alpha}$ given in (\ref{common}) is intuitive because it mirrors (\ref{icc}), it will soon prove useful to have $\tilde{\alpha}$ expressed as
\[
\tilde{\alpha}=\frac{MSA/MSE-1}{MSA/MSE+n-1}.
\]
Some investigators have referred to $MSA/MSE$ as the variance ratio statistic, where the variance ratio in question is $\gamma=\sigma_\tau^2/\sigma_\epsilon^2$. But $MSA/MSE$ is not an estimator of $\gamma$. Rather, $MSA/MSE$ is an (positively biased) estimator of $\theta=n\gamma+1$. The relationships among $\alpha$, $\gamma$, and $\theta$ can be written as
\[
\alpha=\frac{\theta-1}{\theta+n-1}=\frac{(n\gamma+1)-1}{(n\gamma+1)+n-1}=\frac{n\gamma}{n\gamma+n}=\frac{\gamma}{\gamma+1}.
\]

\subsection{A variant of the analytical estimator}

The estimator that is the focus of this section is a variant of the analytical estimator. This variant was introduced by \citet{bcicc} for the express purpose of producing a reduced-bias estimator of $\alpha$ (see next section).

\citet{bcicc} present their variant (for a balanced design) as
\begin{align}
\label{agamma}
\vv{\alpha}&=\frac{\vv{\gamma}}{1+\vv{\gamma}},
\end{align}
where
\[
\vv{\gamma}=\frac{\{N-a-2\}SSA/SSE-(a-1)}{n(a-1)}
\]
is an unbiased estimator of the variance ratio $\gamma$. To see how $\vv{\alpha}$ is a variant of $\tilde{\alpha}$, observe that \citet{bcicc} estimate $\theta$ (unbiasedly) as
\[
\vv{\theta}=n\vv{\gamma}+1=\frac{MSA}{SSE/(N-a-2)},
\]
in contrast to
\[
\tilde{\theta}=\frac{MSA}{MSE}=\frac{MSA}{SSE/(N-a)}
\]
for the analytical estimator.

Clearly, $\vv{\theta}$ is strictly smaller than $\tilde{\theta}$, which implies that $\vv{\alpha}$ is strictly smaller than $\tilde{\alpha}$. Thus, since $\tilde{\alpha}$ is negatively biased, $\vv{\alpha}$ exacerbates the bias of $\tilde{\alpha}$, and so $\vv{\alpha}$ is not a compelling estimator of $\alpha$ in its own right. \citet{bcicc} simply use $\vv{\alpha}$ as a starting point in their effort to formulate an estimator of $\alpha$ that has a smaller bias than the analytical estimator $\tilde{\alpha}$.

Because characterization of $\vv{\alpha}$'s bias employs second-order Taylor approximations of $\log\vv{\gamma}$ and $\log(1+\vv{\gamma})$, the bias correction relies on the variance of $\vv{\gamma}$, which is equal to  \citep{bcicc}
\begin{align}
\label{vargam}
\var\vv{\gamma}=\frac{N-a-2}{n^2(a-1)}\left\{\frac{a+1}{N-a-4}-\frac{a-1}{N-a-2}\right\}(n\gamma+1)^2.
\end{align}
One must estimate this variance since it is a function of $\gamma$. \citet{bcicc} recommend the plug-in estimator that replaces $\theta=n\gamma+1$ in (\ref{vargam}) with $\vv{\theta}=n\vv{\gamma}+1$. Of course another possibility is to substitute $\tilde{\theta}=MSA/MSE$ for $\theta$.

\subsection{A bias-corrected estimator}

\citet{bcicc} characterized the bias of $\vv{\alpha}$ and proposed two bias-corrected estimators, which I will denote as $\vv{\alpha}_\text{bc1}$ and $\vv{\alpha}_\text{bc2}$. To reveal the bias, first write second-order Taylor approximations of $\log\vv{\gamma}$ and $\log(1+\vv{\gamma})$ as
\[
\log\vv{\gamma}\approx\log\gamma+\frac{1}{\gamma}(\vv{\gamma}-\gamma)-\frac{1}{2\gamma^2}(\vv{\gamma}-\gamma)^2
\]
and
\[
\log(\vv{\gamma}+1)\approx\log(\gamma+1)+\frac{1}{\gamma+1}(\vv{\gamma}-\gamma)-\frac{1}{2(\gamma+1)^2}(\vv{\gamma}-\gamma)^2
\]
for $\gamma$ near $\vv{\gamma}$. This implies that
\[
\e(\log\vv{\alpha}-\log\alpha)\approx-\frac{1}{2}\left\{\frac{1}{\gamma^2}-\frac{1}{(\gamma+1)^2}\right\}\var\vv{\gamma}.
\]
Then produce a bias-corrected estimator of $\log\alpha$ as
\[
\vv{\log\alpha}_\text{bc}=\log\vv{\alpha}+\frac{1}{2}\left\{\frac{1}{\vv{\gamma}^{\,2}}-\frac{1}{(\vv{\gamma}+1)^2}\right\}\widehat{\var}\vv{\gamma}.
\]
Here $\widehat{\var}\vv{\gamma}$ is an estimator of $\var\vv{\gamma}$. Finally, map back to the original scale to obtain
\begin{align}
\label{abc1}
\vec{\alpha}_\text{bc1}&=\vec{\alpha}\exp\left[\frac{1}{2}\left\{\frac{1}{\vec{\gamma}^{\,2}}-\frac{1}{(\vec{\gamma}+1)^2}\right\}\widehat{\var}\vec{\gamma}\right].
\end{align}

We can already see that this estimator will tend to perform poorly when agreement is low. This is because the dilation factor in (\ref{abc1}) goes to $\infty$ as $\gamma$ goes to 0. When the sample size is smaller and agreement is low, $\vec{\alpha}_\text{bc1}$ is even challenging to evaluate by simulation because the estimator can take any value in $\reals\cup\{-\infty,\infty\}$ and has very heavy tails. Since this estimator's drawbacks do not vanish until the sample size is large, I will not consider $\vec{\alpha}_\text{bc1}$ further.

\subsection{A second bias-corrected estimator}

\citet{bcicc} noted the above mentioned deficiency of $\vec{\alpha}_\text{bc1}$ and recommended an alternative estimator (based on Taylor expansion of $1-\alpha$) in the case of small $\gamma$. Said alternative is given by
\begin{align}
\label{abc2}
\vec{\alpha}_\text{bc2}&=1-(1-\vec{\alpha})\exp\left\{-\frac{1}{2(\vec{\gamma}+1)^2}\widehat{\var}\vec{\gamma}\right\}.
\end{align}
The sensible contraction factor for $\vec{\alpha}_\text{bc2}$ allows $\vec{\alpha}_\text{bc2}$ to perform much better than $\vec{\alpha}_\text{bc1}$ for small values of $\gamma$.

\subsection{Choosing the best point estimator}

\begin{figure}[ht]
   \begin{center}
   \begin{tabular}{ccc}
   \includegraphics[scale=.4]{long.jpg}\;\;\; $16\times 4$ & \includegraphics[scale=.4]{square.jpg}\;\;\; $8\times 8$ & \includegraphics[scale=.4]{short.jpg}\;\;\; $4\times 16$\\
   \includegraphics[scale=.23]{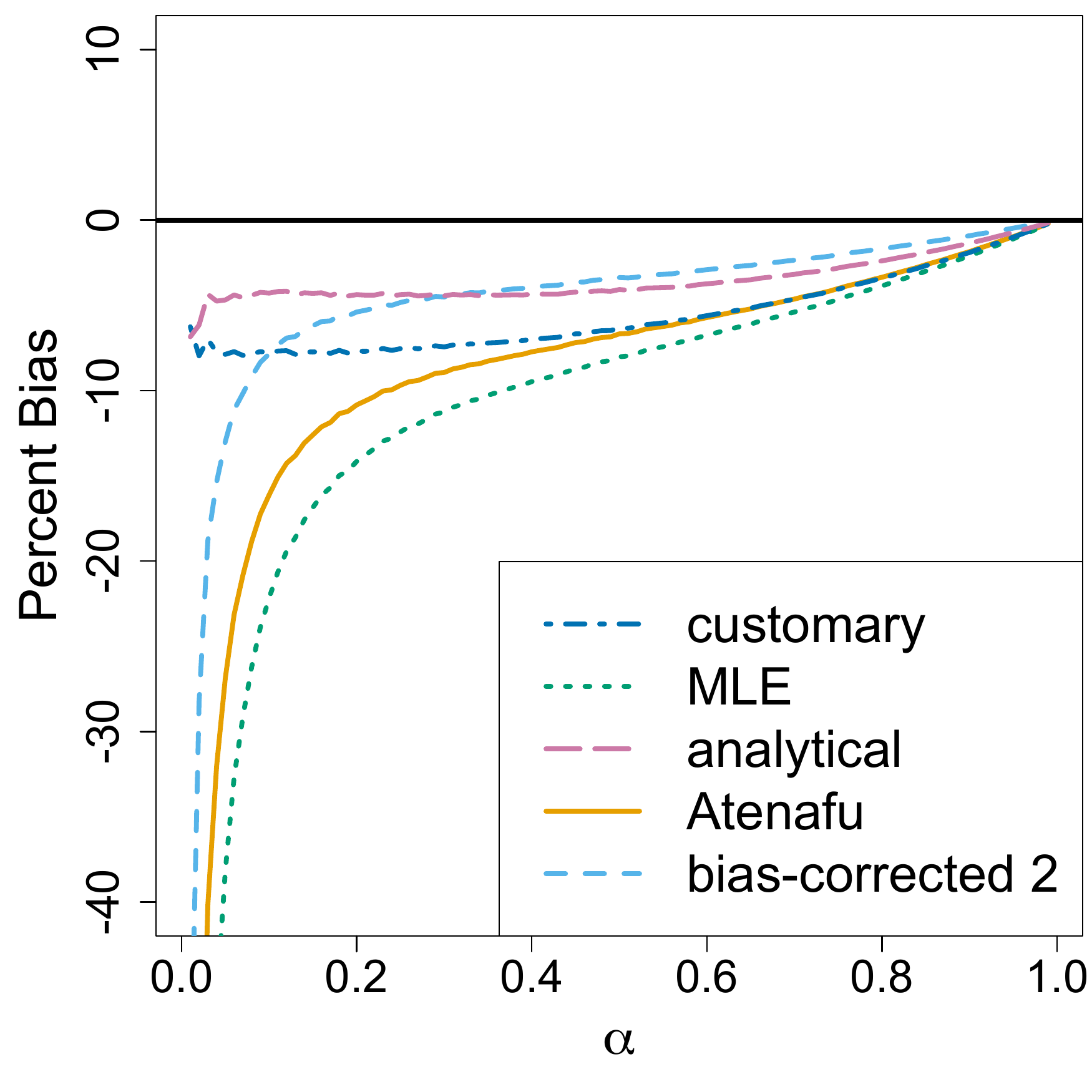} & \includegraphics[scale=.23]{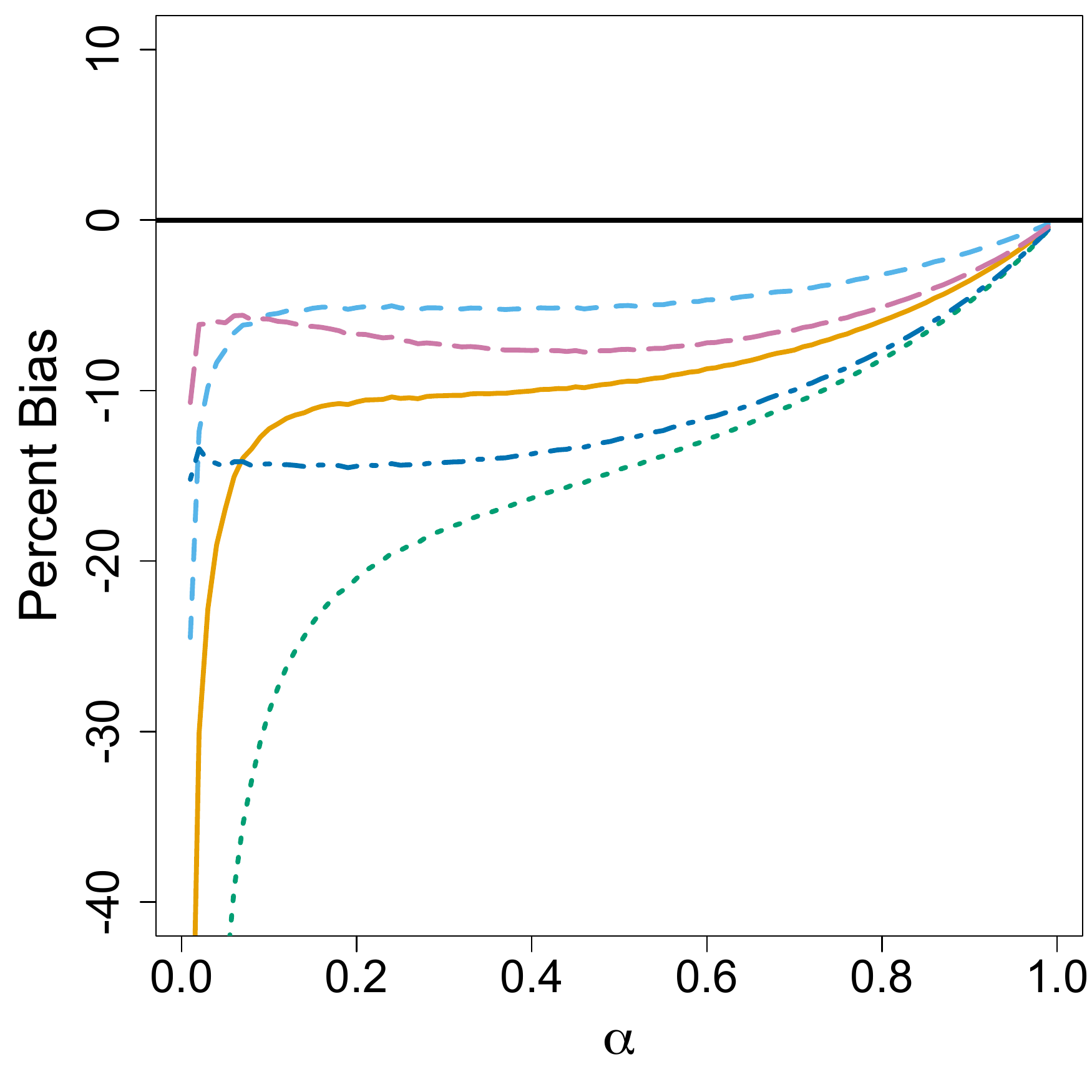} & \includegraphics[scale=.23]{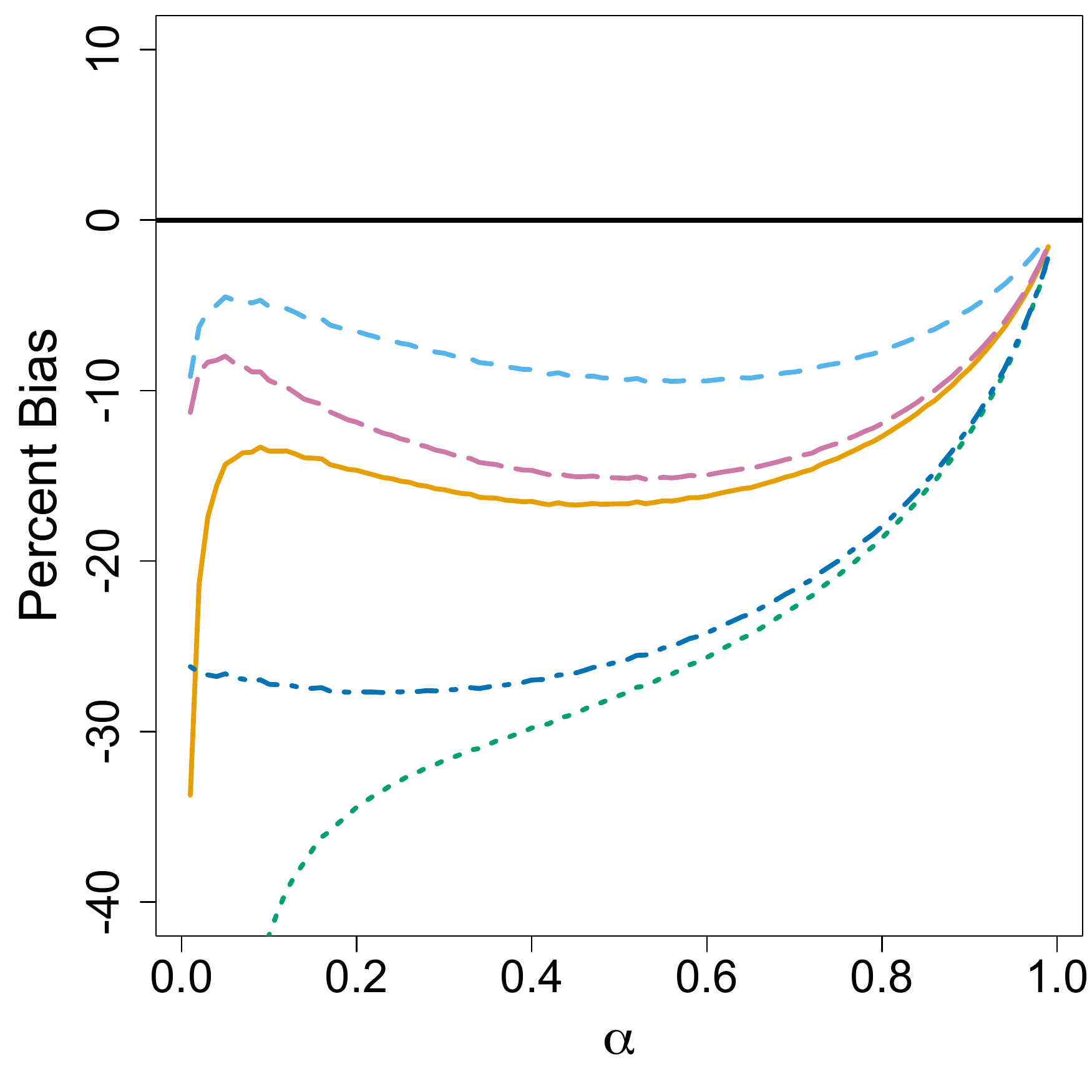} \\
   \includegraphics[scale=.23]{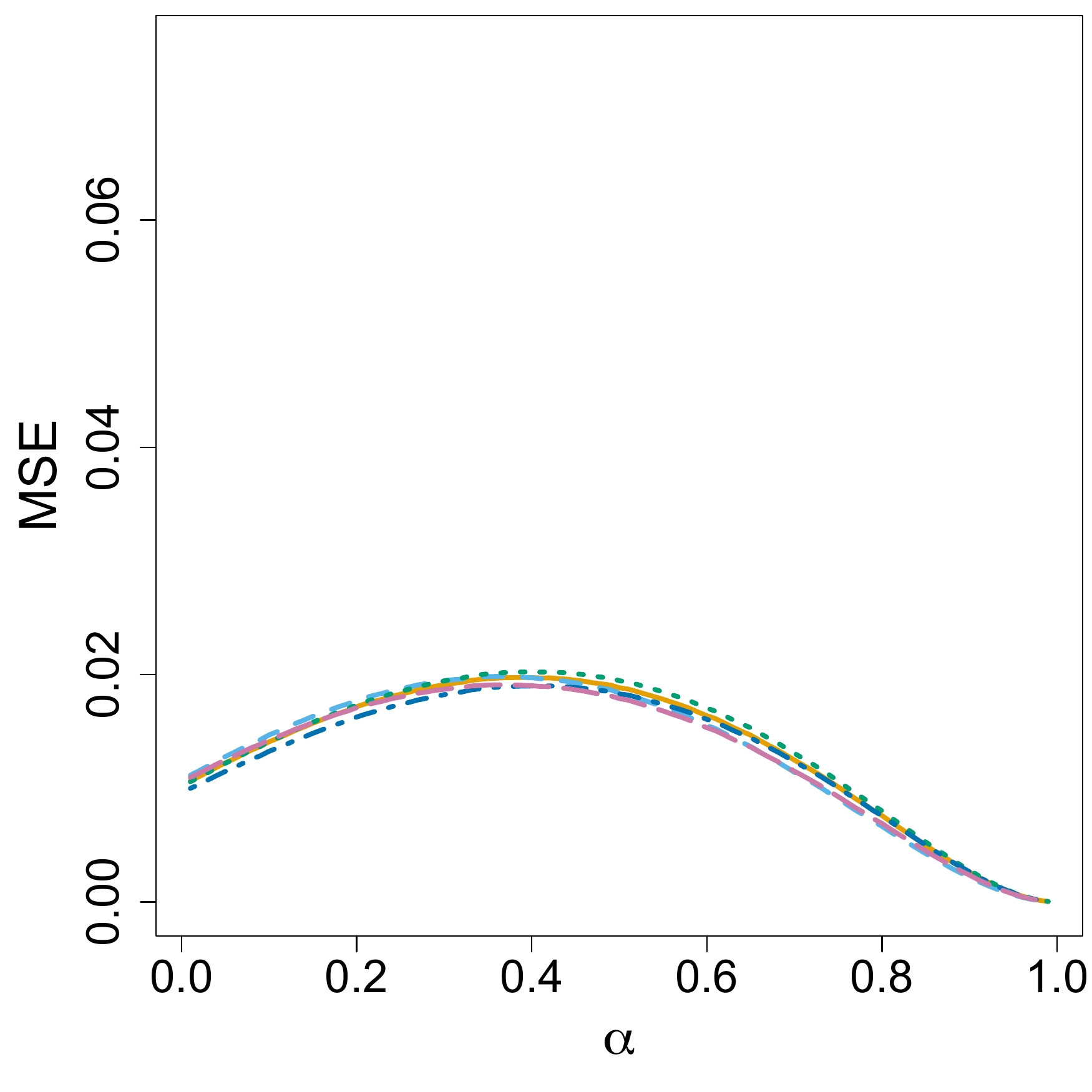} & \includegraphics[scale=.23]{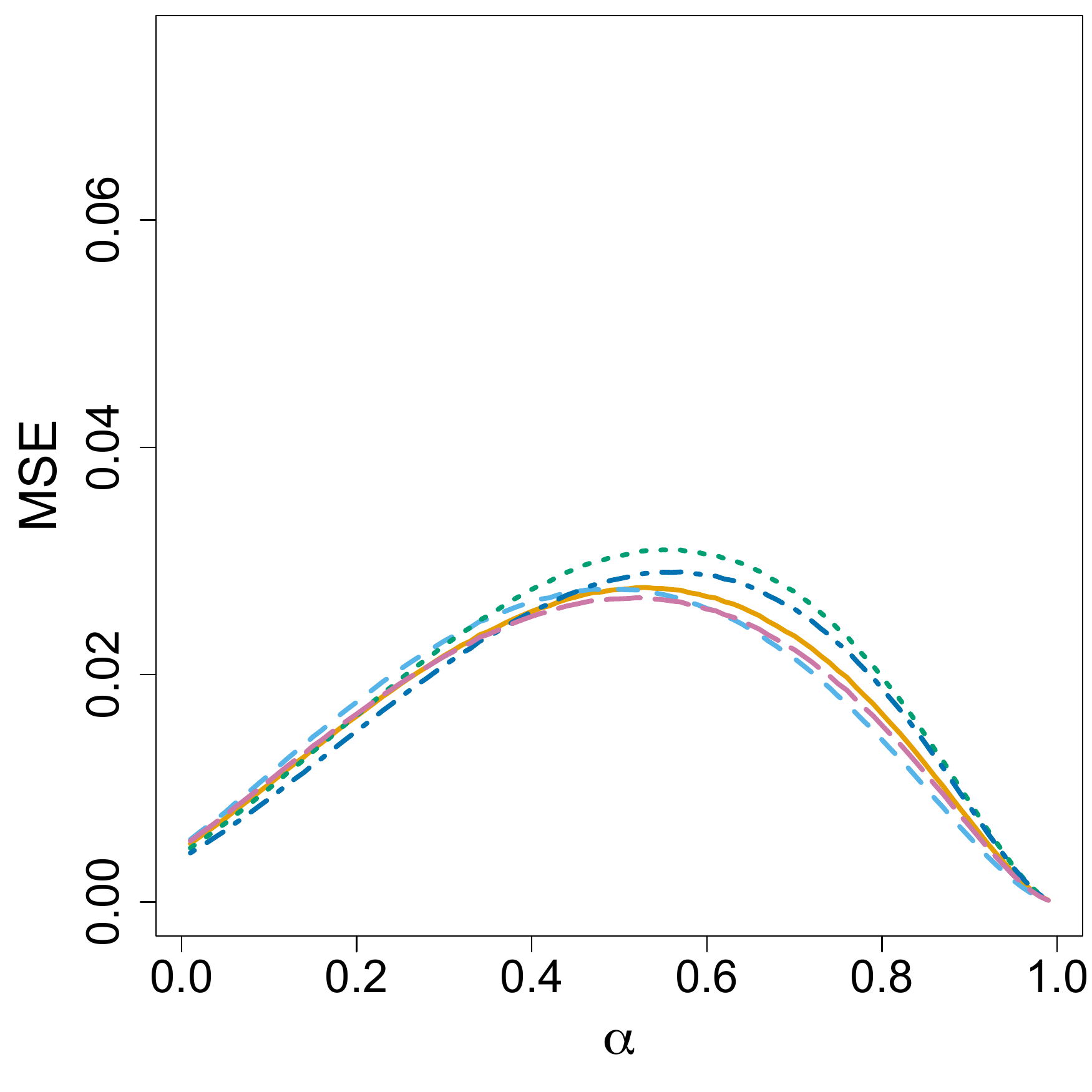} & \includegraphics[scale=.23]{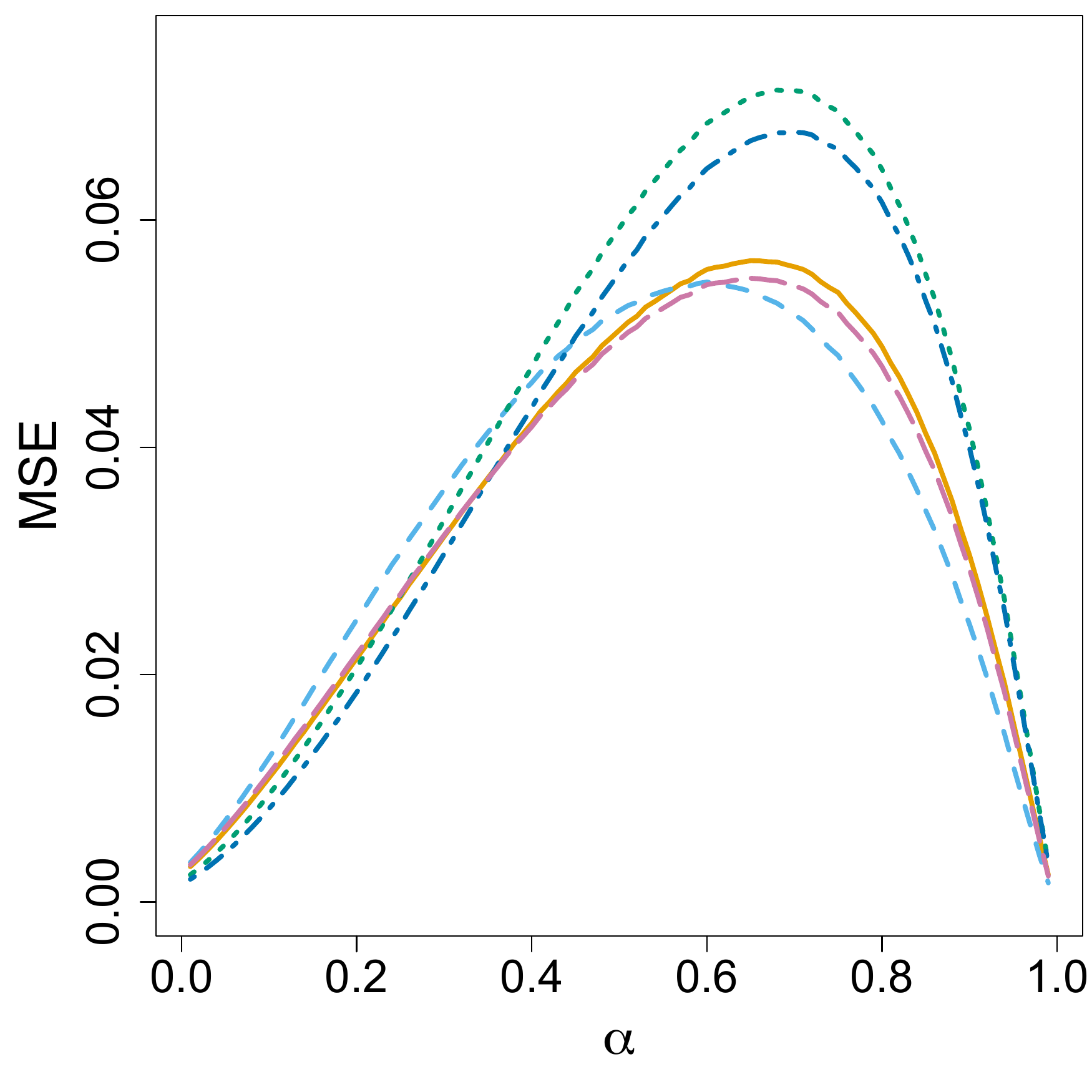} 
   \end{tabular}
   \end{center}
   \caption{Percent bias and mean squared error for the five point estimators.}
   \label{estimators}
\end{figure}

I compared the five point estimators to one another by way of an extensive simulation study. Specifically, for each value of $\alpha$ in a fine grid with step size 0.01, I simulated 1,000,000 datasets and computed each of the five estimates for each dataset. Graphical results for percent bias and mean squared error are shown in Figure~\ref{estimators}.

The contest is clearly between the analytical estimator and the second bias-corrected estimator of \citet{bcicc}. For tall datasets, the analytical estimator performs about as well as the bias-corrected estimator for $\alpha$ values between approximately 0.2 and 1. As $\alpha$ descends toward 0, the analytical estimator exhibits less and less bias relative to the bias-corrected estimator. The two estimators are nearly indistinguishable with respect to mean squared error.

For the square-matrix scenario, the bias-corrected estimator exhibits less bias than the analytical estimator over nearly the entire interval, although the difference is small. And once again the estimators' mean squared errors are close for all values of $\alpha$.

For the short-matrix scenario, the bias-corrected estimator offers substantially less bias than the analytical estimator. As for mean squared error, we can see a noticeable, but not too large, difference in this scenario. For small to moderate agreement the analytical estimator is more accurate. For moderate to strong agreement the bias-corrected estimator is more accurate.

In summary, the analytical estimator is better for tall data matrices, the two estimators perform comparably for square matrices, and the bias-corrected estimator is better for short matrices. Either the analytical estimator or the bias-corrected estimator is a better choice than the customary estimator, the MLE, or the analytical variant proposed by \citet{bcicc}. Of course these differences in performance disappear as the sample size grows large. For smaller sample sizes, though, making an educated choice can have a meaningful impact on the quality of inference.

\section{Jackknife-based interval estimation for the analytical point estimator}

Whether one chooses the analytical estimator or the bias-corrected estimator, it is desirable to obtain confidence intervals having the desired rates of coverage. Above I argued and verified by simulation that bootstrapping based on resampling the rows of the data matrix yields unacceptably low coverage rates. In this section I will develop an effective jackknife method for interval estimation that can be applied to either estimator. I will present the method for $\tilde{\alpha}$.

The jackknife variance estimator for $\tilde{\alpha}$ is based on $\log(MSA/MSE)=\log(\tilde{\theta})$ because log-transforming $MSA/MSE$ normalizes the estimator and stabilizes its variance. This permits one to compute a good-performing Student's $t$-based confidence interval for $\log(\theta)$, the endpoints of which are then transformed to the scale of $\alpha$ using $\alpha=(\theta-1)/(\theta+n-1)$.

Let $\eta=\log(\theta)$. Then compute $a$ pseudovalues
\[
\dot{\eta}_i=a\tilde{\eta}-(a-1)\eta_{-i},
\]
where $\tilde{\eta}=\log(\tilde{\theta})$ and $\eta_{-i}$ is equal to the analytical estimate of $\eta$ for the data matrix with row $i$ left out. Then the jackknife estimator of variance is $V_\text{jack}=S^2/a$, where $S^2$ is the sample variance of the pseudovalues:
\[
S^2=\frac{\sum_{i=1}^a(\dot{\eta}_i-\frac{1}{a}\sum_{i=1}^a\dot{\eta}_i)^2}{a-1}.
\]
The statistic
\[
T=\frac{\tilde{\eta}-\eta}{\sqrt{V_\text{jack}}}
\]
is approximately $t$ distributed, and so an estimated $(1-\delta)100\%$ confidence interval for $\eta$ is given by
\[
(L_\eta=\tilde{\eta}-t_\nu^{1-\delta/2}\sqrt{V_\text{jack}},\;U_\eta=\tilde{\eta}+t_\nu^{1-\delta/2}\sqrt{V_\text{jack}}),
\]
where $t_\nu^{1-\delta/2}$ denotes the $1-\delta/2$ quantile of Students's $t$ distribution with $\nu$ degrees of freedom. The corresponding interval for $\alpha$ is
\[
\left(L=\frac{\exp(L_\eta)-1}{\exp(L_\eta)+n-1},\;U=\frac{\exp(U_\eta)-1}{\exp(U_\eta)+n-1}\right).
\]

Given that $S^2$ is the sample variance for a sample of size $a$, one might suspect that $\nu=a-1$. Alas, $\nu$ is complicated and so must be estimated or simply set to $a-1$. I investigated the double-jackknife estimation approach recommended by \citet{hinkley1977jackknife}, wherein $\nu$ is estimated as
\[
\tilde{\nu}=\frac{2V_\text{jack}^2}{K},
\]
where
\[
K=\frac{\sum_{i=1}^a(\dot{\eta}-\frac{1}{a}\sum_{i=1}^a\dot{\eta}_i)^4}{a(a-1)(a-2)^2}-\frac{aV_\text{jack}^2}{(a-2)^2},
\]
but found that using $\tilde{\nu}$ does not improve on using $a-1$ in this setting because $\tilde{\nu}$ is too variable for smaller samples. Luckily, assuming $a-1$ degrees of freedom yields coverage rates close to $(1-\delta)100\%$ because $\nu$ is approximately equal to $a-1$ for small values of $a$, and then $\nu$ and $a$ diverge in a complicated way as $a$ increases.

\begin{figure}[ht]
   \begin{center}
   \begin{tabular}{ccc}
   \includegraphics[scale=.4]{long.jpg}\;\;\; $16\times 4$ & \includegraphics[scale=.4]{square.jpg}\;\;\; $8\times 8$ & \includegraphics[scale=.4]{short.jpg}\;\;\; $4\times 16$\\
   \includegraphics[scale=.23]{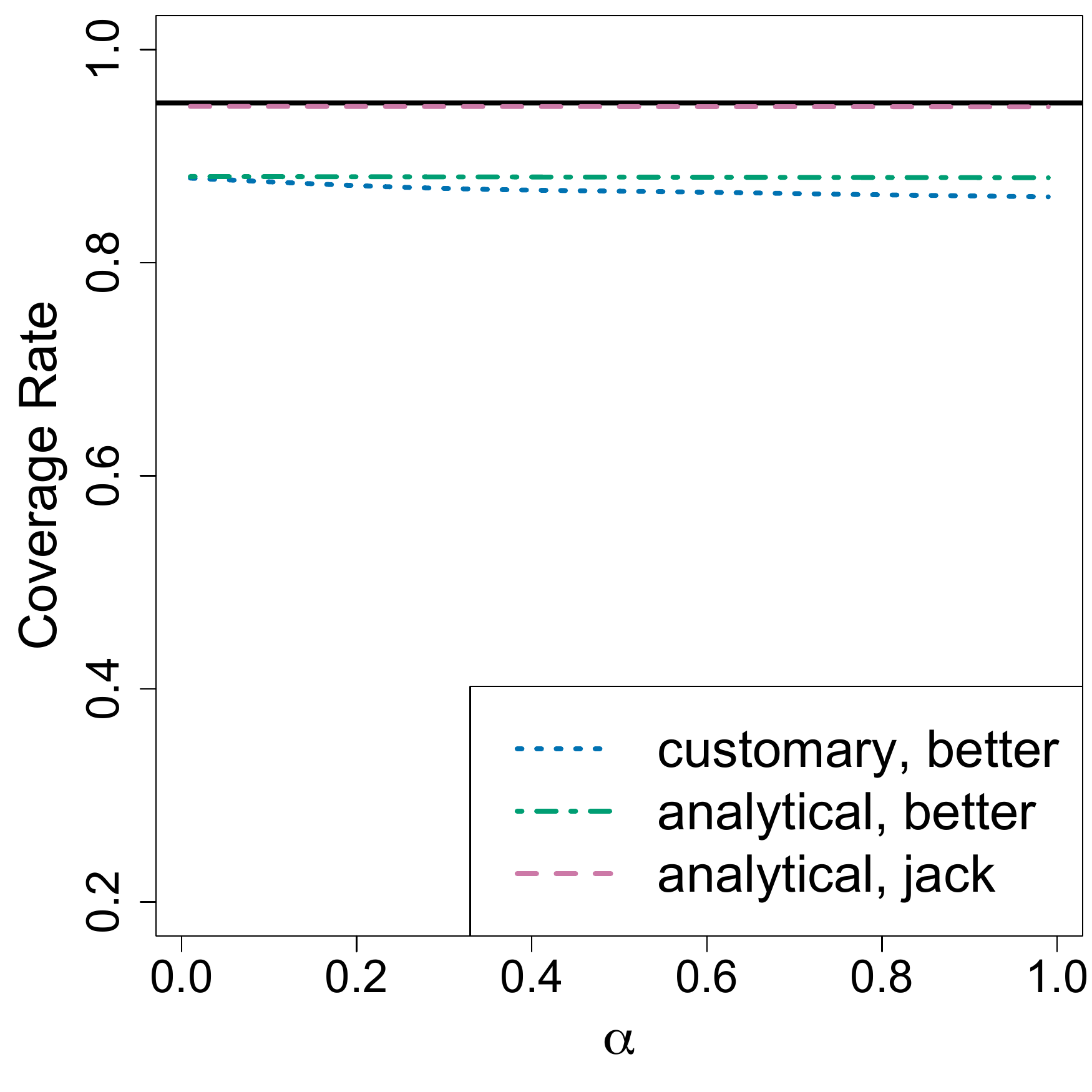} & \includegraphics[scale=.23]{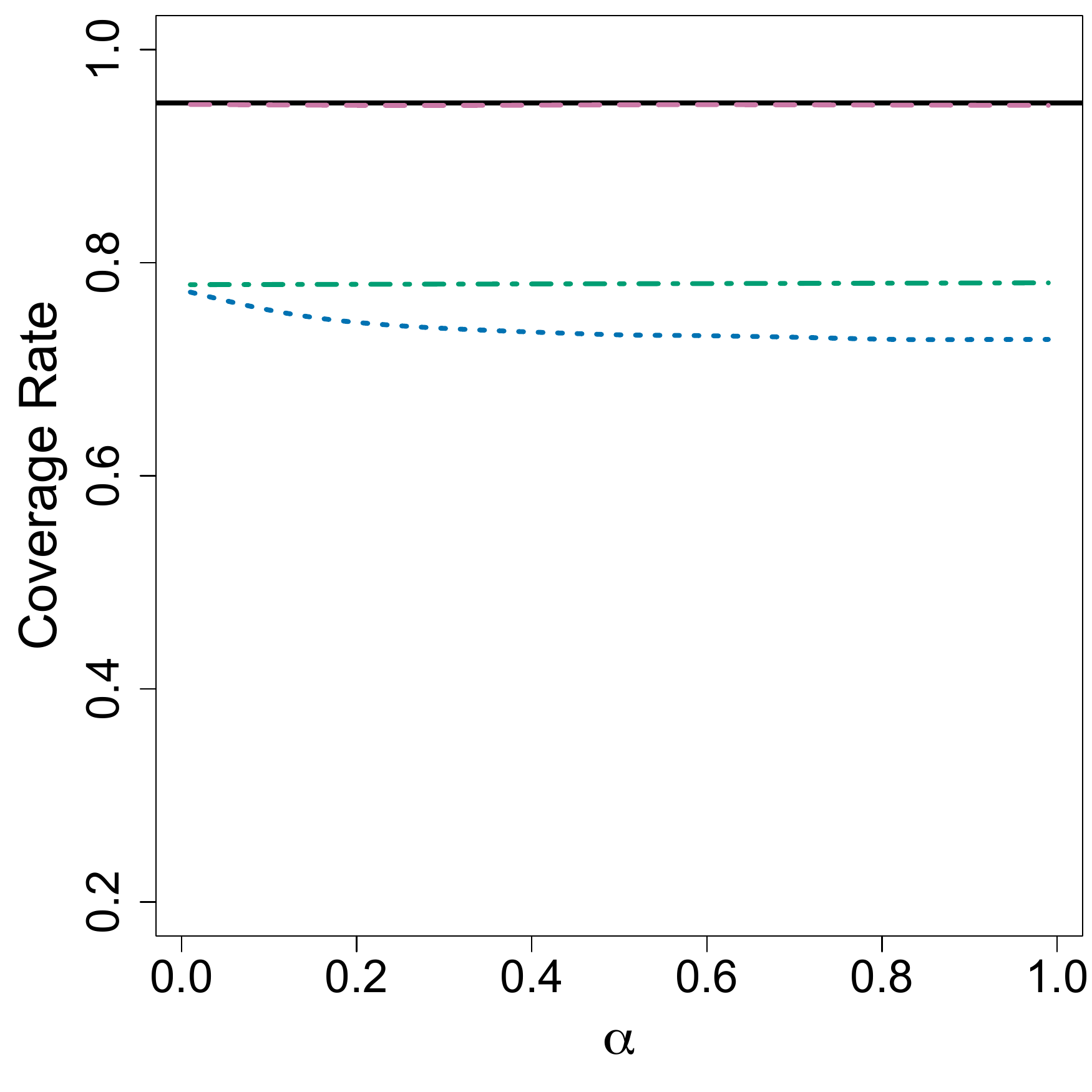} & \includegraphics[scale=.23]{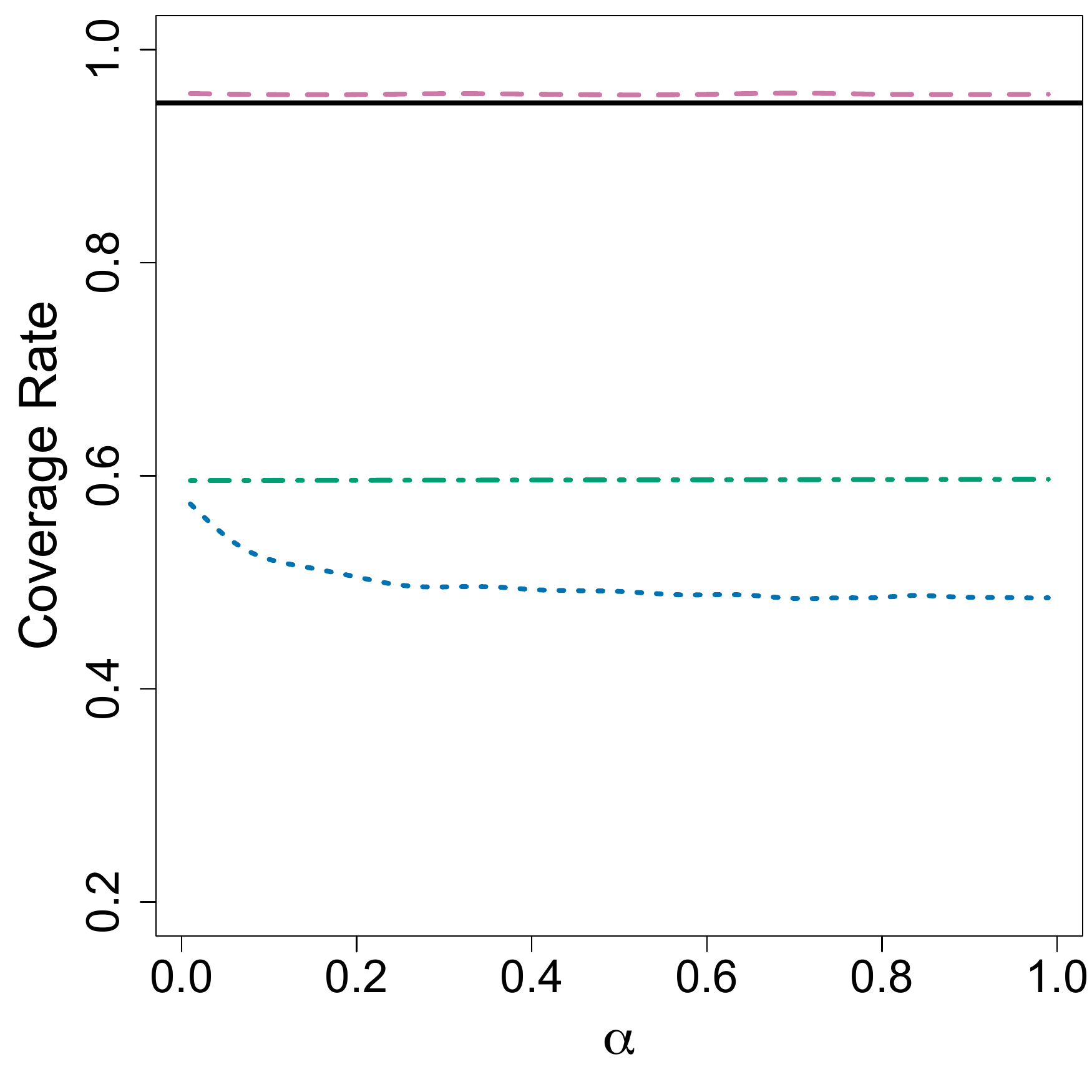} \\
   \end{tabular}
   \end{center}
   \caption{Coverage rates of 95\% intervals for the customary Krippendorff's $\alpha$ estimator with improved bootstrap procedure (blue, dotted), the analytical estimator with improved bootstrap procedure (green, dash-dot), the analytical estimator with jackknife variance estimation (pink, dashed).}
   \label{ci3}
\end{figure}

Simulation results are shown in Figure~\ref{ci3}. Each panel shows coverage rates for the customary point estimator $\hat{\alpha}$ with the improved bootstrap procedure described in Section~\ref{salvage}, the analytical estimator $\tilde{\alpha}$ with improved bootstrap, and the analytical estimator with jackknife variance estimation. We see that the coverage rates are very close to 95\% for the latter method while the first two methods yield poor coverage rates.

\section{Computing issues}

The time complexities of the four approaches to interval estimation are given in Table~\ref{runningtimes}, where $b$ is the bootstrap sample size, $a$ is the number of units, and $n$ is the number of coders. These rates are for the general, i.e., nonparametric, version of Krippendorff's $\alpha$, which is arrived at by applying the identity
\[
\sum_{i=1}^m(x_i-\bar{x}_\bull)^2=\frac{1}{2m}\sum_{i=1}^m\sum_{j=1}^m(x_i-x_j)^2
\]
to the usual definitions of $SSE$ and $SST_c$ and then allowing for other (perhaps even user-defined) distance functions in addition to squared Euclidean distance $d^2(x_i,x_j)=(x_i-x_j)^2$ \citep{kripppkg}.

We see that the jackknife procedure's growth rate is cubic in $a$ while the other procedures have quadratic running times. These fast growth rates can make analyses of larger datasets quite burdensome or even infeasible. This is a minor problem, though, since for large datasets even the customary bootstrap procedure provides high-quality inference. Still, it is desirable to carefully consider how one might speed computation of $SST_c$, which is $\Theta(a^2n^2)$ and must be computed $b$ times for the improved bootstrap procedure and $a$ times for the jackknife method.

\begin{table}[ht]
\caption{Running times for the four approaches to interval estimation.}
\label{runningtimes}
   \begin{center}
   \begin{tabular}{ll}
   Method & Time Complexity\\\hline
   $\hat{\alpha}$ with customary bootstrap &  $\Theta(a^2n^2)$\\
   $\hat{\alpha}$ with improved bootstrap &  $\Theta(ba^2n^2)$\\
   $\tilde{\alpha}$ with improved bootstrap &  $\Theta(ba^2n^2)$ \\
   $\tilde{\alpha}$ with jackknife variance estimation &  $\Theta(a^3n^2)$
    \end{tabular}
    \end{center}
\end{table}

To see why computation of $SST_c$ is onerous in the general case, consider the nonparametric version of $SST_c$:
\[
SST_c=\frac{1}{2an}\sum_{i=1}^a\sum_{j=1}^n\sum_{k=1}^a\sum_{l=1}^nd^2(Y_{ij},Y_{kl}),
\]
where $d^2$ is the chosen distance function. Clearly, this entails computing the distance between every pair of outcomes in the dataset. This computational load can be eased considerably by storing the distances in an $an\times an$ lookup table during the first pass over the data, and then using the lookup table to compute $(SST_c)_k^*\;\;(k=1,\dots,b)$ for the bootstrap or $(SST_c)_{-i}\;\;(i=1,\dots,a)$ for the jackknife. Employing a lookup table can be hundreds of times faster than computing $SST_c$ from scratch during each iteration.

\section{Additional analyses of simulated data}
\label{simstudy}

In addition to carrying out the simulation experiments already described, I investigated by simulation the behaviors of the customary and analytical estimators when (1) the outcomes were Gaussian and the design was unbalanced (20\% missing at random), (2) the outcomes were continuous but the unit effects were Student's $t$ distributed with four degrees of freedom, and (3) when the outcomes were categorical.

For the unbalanced Gaussian scenario and the $t$ scenario, $\alpha$ is still a well-defined population parameter, and so I was able to measure bias, mean squared error, and coverage rates, as above. I have omitted those results because they are quite similar to the results I presented above for Gaussian unit effects in a balanced design.

Investigating the behavior of $\hat{\alpha}$ and $\tilde{\alpha}$ for categorical outcomes is a rather different matter since the population parameter cannot be defined precisely. This poses two challenges. First, since $\alpha$ does not belong to a known probability model, one must choose a probability model to serve as a sort of proxy. Second, only estimation and variance can be appraised due to the mismatch between the proxy probability model and the unknown true probability model.

A sensible proxy model from which to simulate categorical outcomes is the direct Gaussian copula model with compound symmetry dependence structure and categorical marginal distribution \citep{xue2000multivariate,dtexact}. The generative form of this model is given by
\begin{align*}
\nonumber\bZ &\;\sim\; \textsc{Normal}\{\bzero,\momega(\alpha)\}\\
\nonumber U_{ij} &\;=\; \Phi(Z_{ij})\;\;\;\;\;\;\;\;\;\;\;\;\;\;\;\;\;\;\;\;\;\;\;(i=1,\dots,a;\; j=1,\dots,n)\\
Y_{ij} &\;=\; F^{-1}(U_{ij}\mid\bpi),
\end{align*}
where $\momega$ is block diagonal with each $n\times n$ block having compound symmetry structure
\[
\momega_i = \bordermatrix{ & 1 & 2 & \dots & n \cr
1 & 1 & \alpha  &\dots & \alpha\cr
2 & \alpha &  1 &  \dots & \alpha\cr
\vdots & \vdots & \vdots  & \ddots  & \vdots\cr
n & \alpha &  \alpha & \dots  & 1
},
\]
$\Phi$ is the standard Gaussian cdf, and $F^{-1}(\cdot\mid\bpi)$ is the quantile function for the categorical distribution having probabilities $\bpi=(\pi_1,\dots,\pi_p)'$. Here $\bU=(U_{11},\dots, U_{an})'$ is a realization of the Gaussian copula indexed by $\momega$, which is to say that the $U_{ij}$ are marginally standard uniform and exhibit the Gaussian correlation structure defined by $\momega$. Since $U_{ij}$ is standard uniform, applying the inverse probability integral transform to $U_{ij}$ in the final stage produces outcome $Y_{ij}$ having the desired categorical marginal distribution $F(\cdot\mid\bpi)$.

\begin{figure}[ht]
   \begin{center}
   \begin{tabular}{ccc}
  \includegraphics[scale=.4]{long.jpg}\;\;\; $16\times 4$ & \includegraphics[scale=.4]{square.jpg}\;\;\; $8\times 8$ & \includegraphics[scale=.4]{short.jpg}\;\;\; $4\times 16$\\
  \includegraphics[scale=.23]{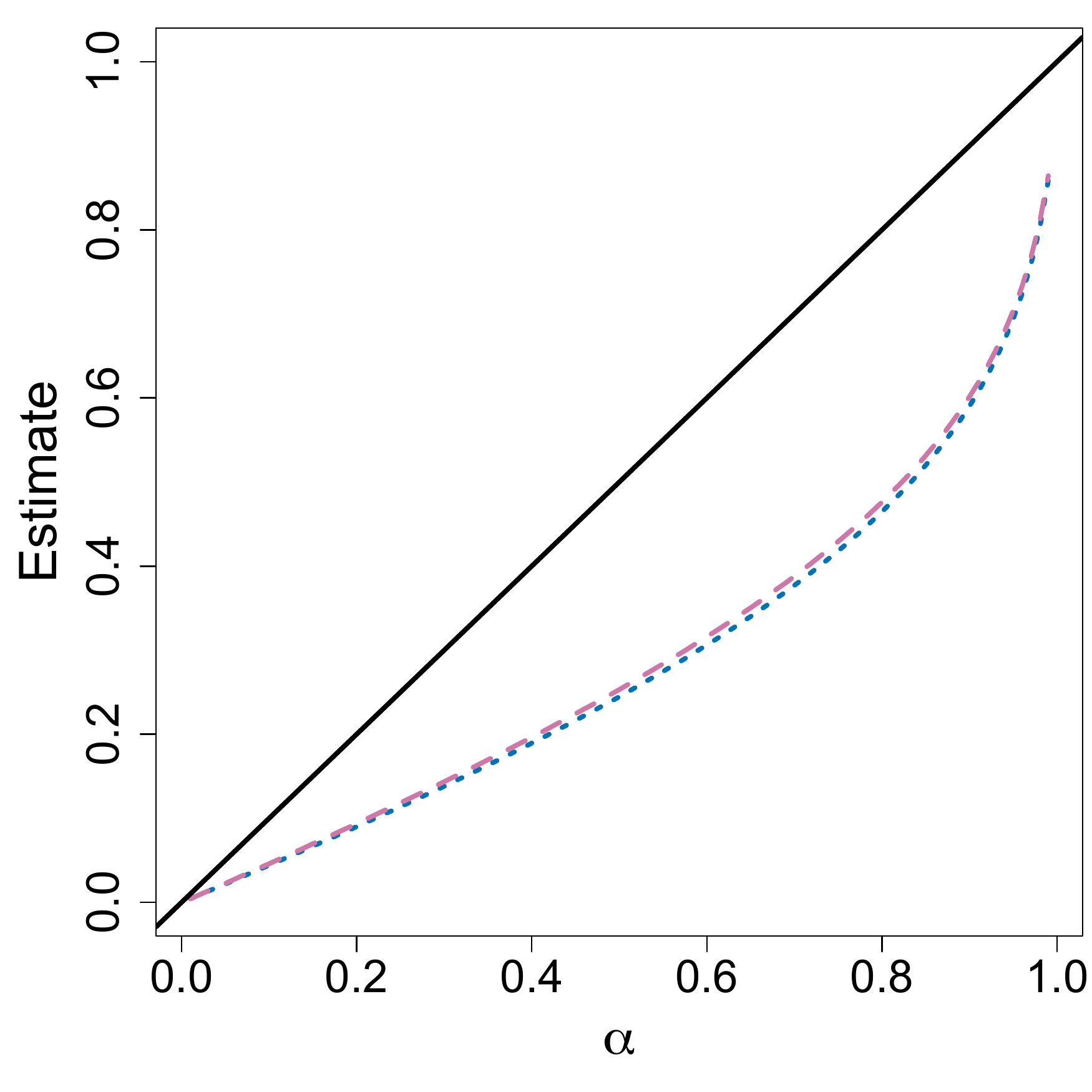} & \includegraphics[scale=.23]{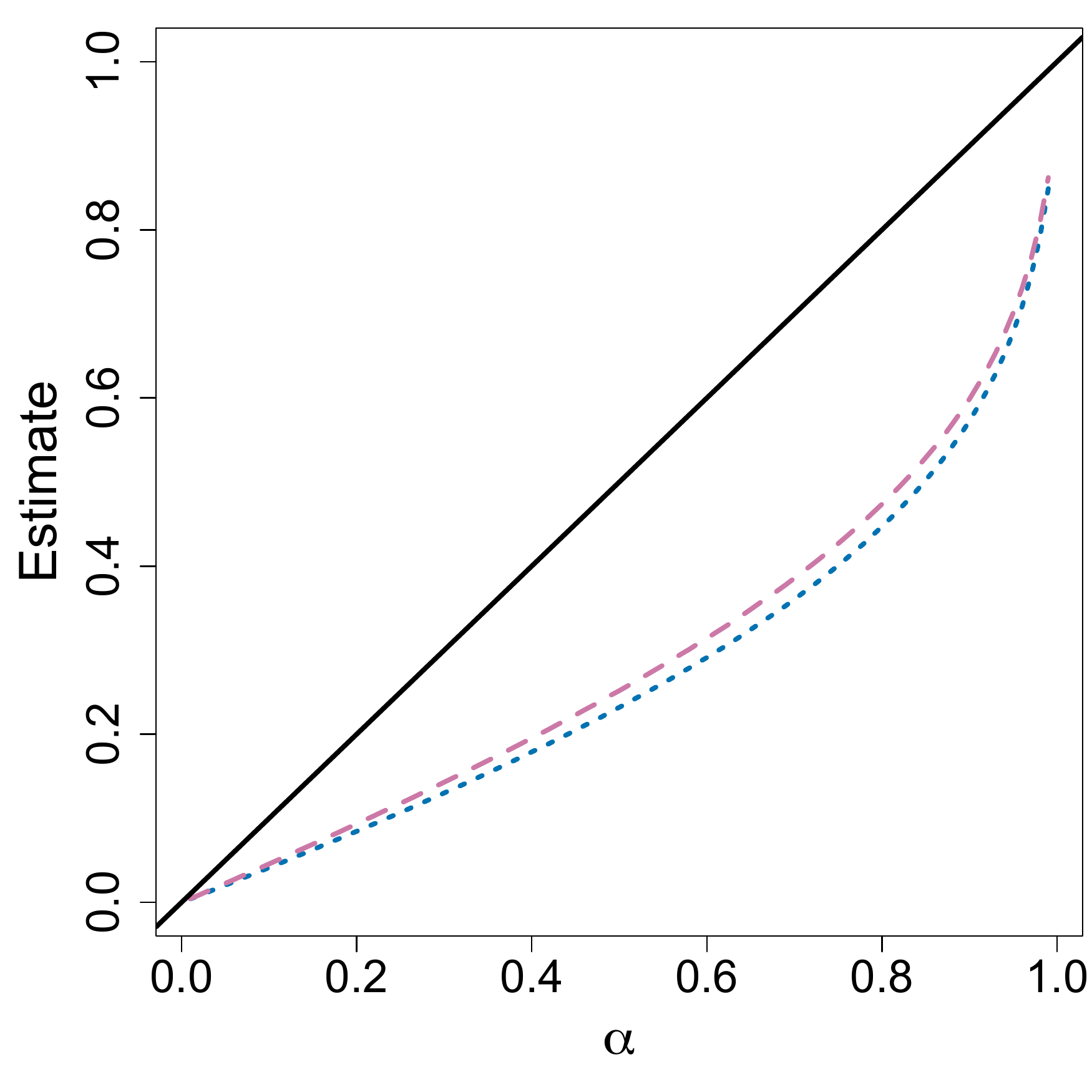} & \includegraphics[scale=.23]{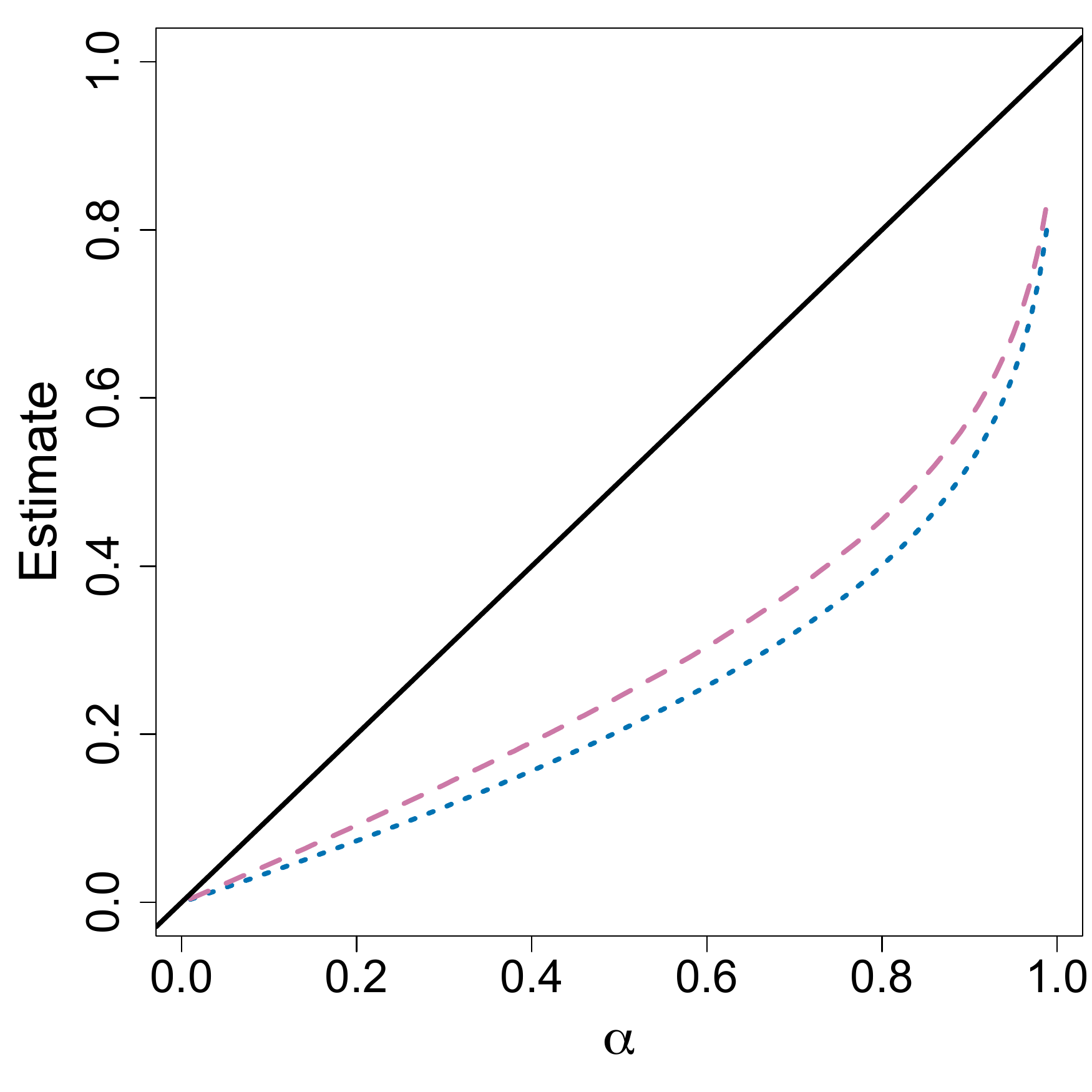} \\
& Categorical &\\
&&\\
 \includegraphics[scale=.23]{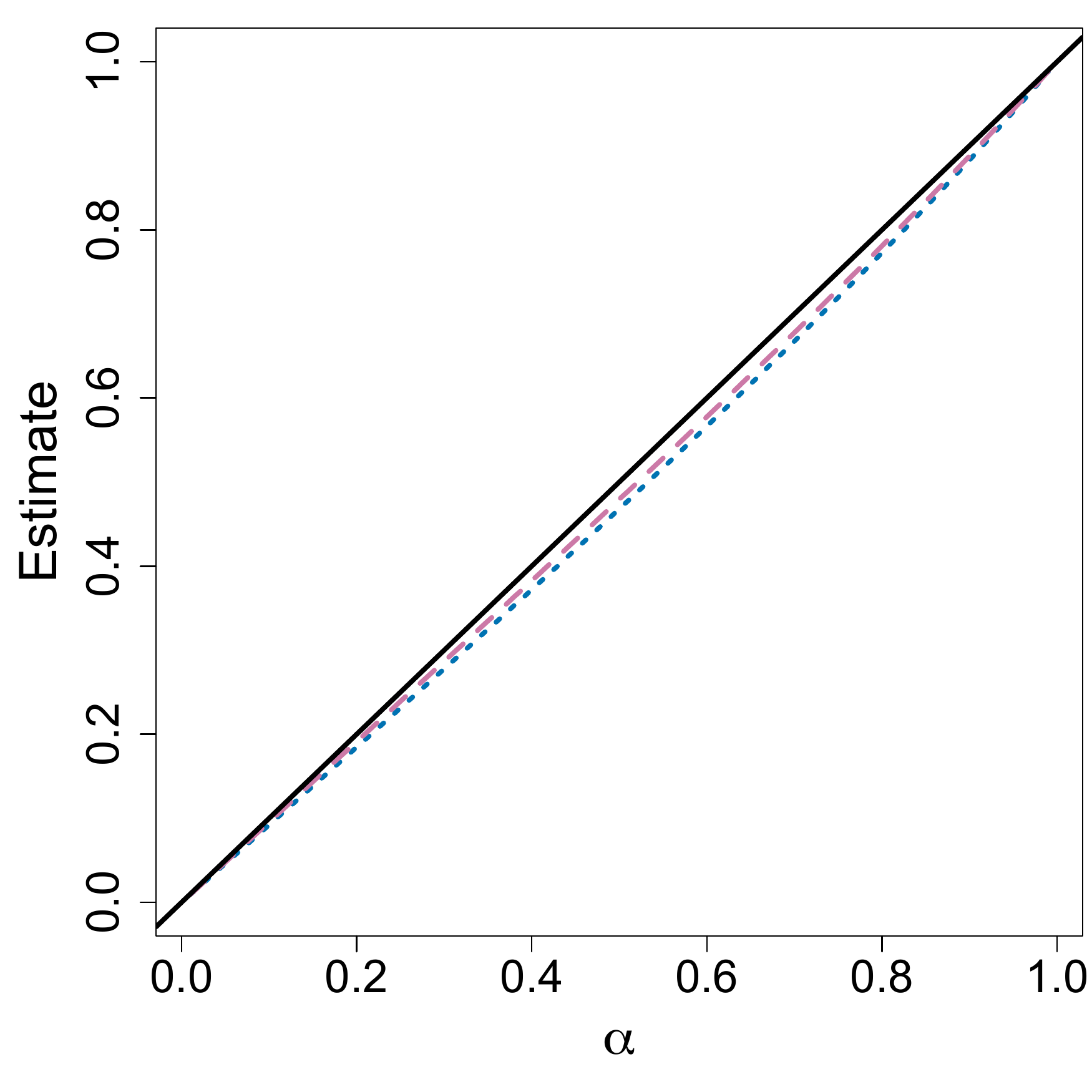} & \includegraphics[scale=.23]{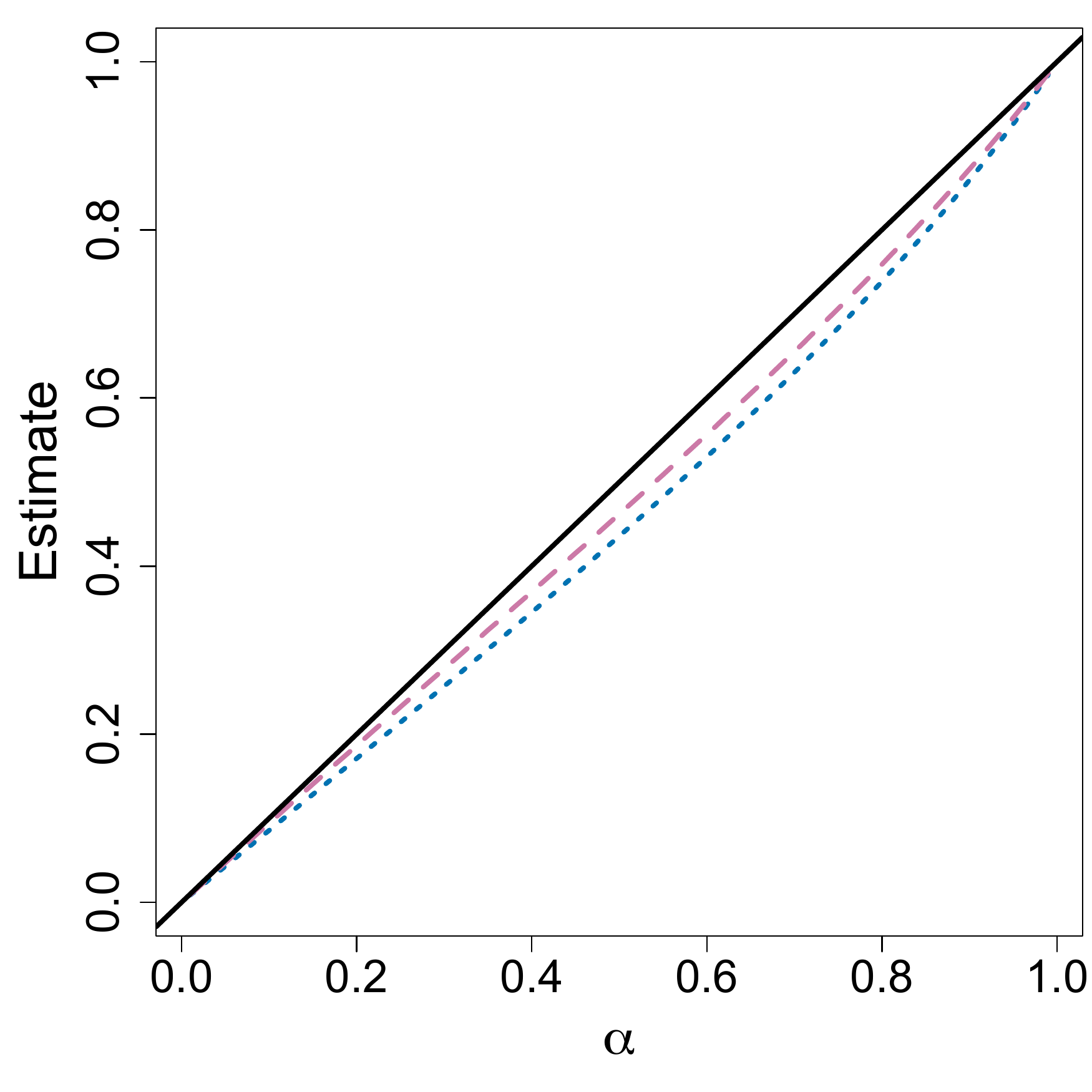} & \includegraphics[scale=.23]{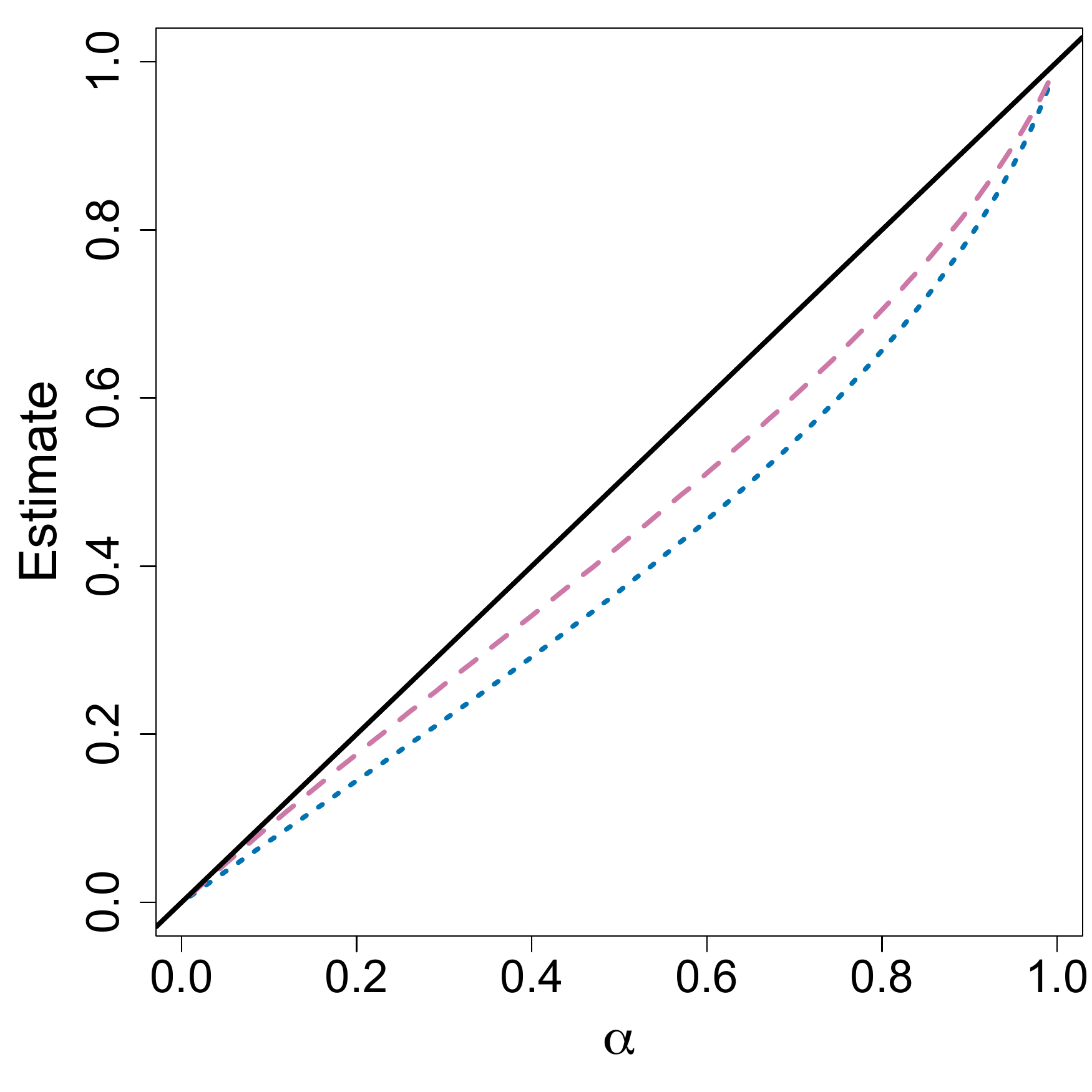} \\
 & Gaussian &\\
 &&\\\hline
 &&\\
\includegraphics[scale=.23]{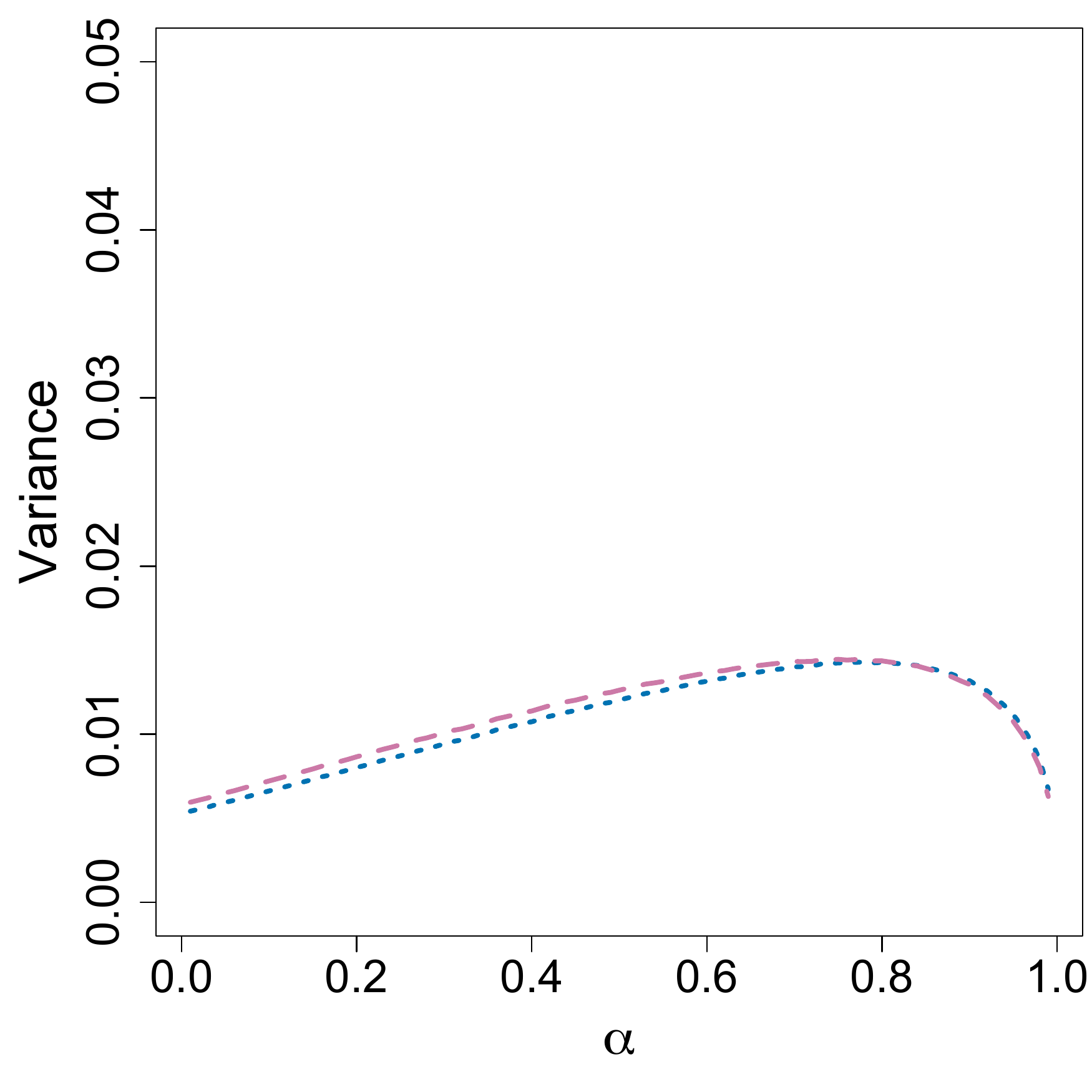} & \includegraphics[scale=.23]{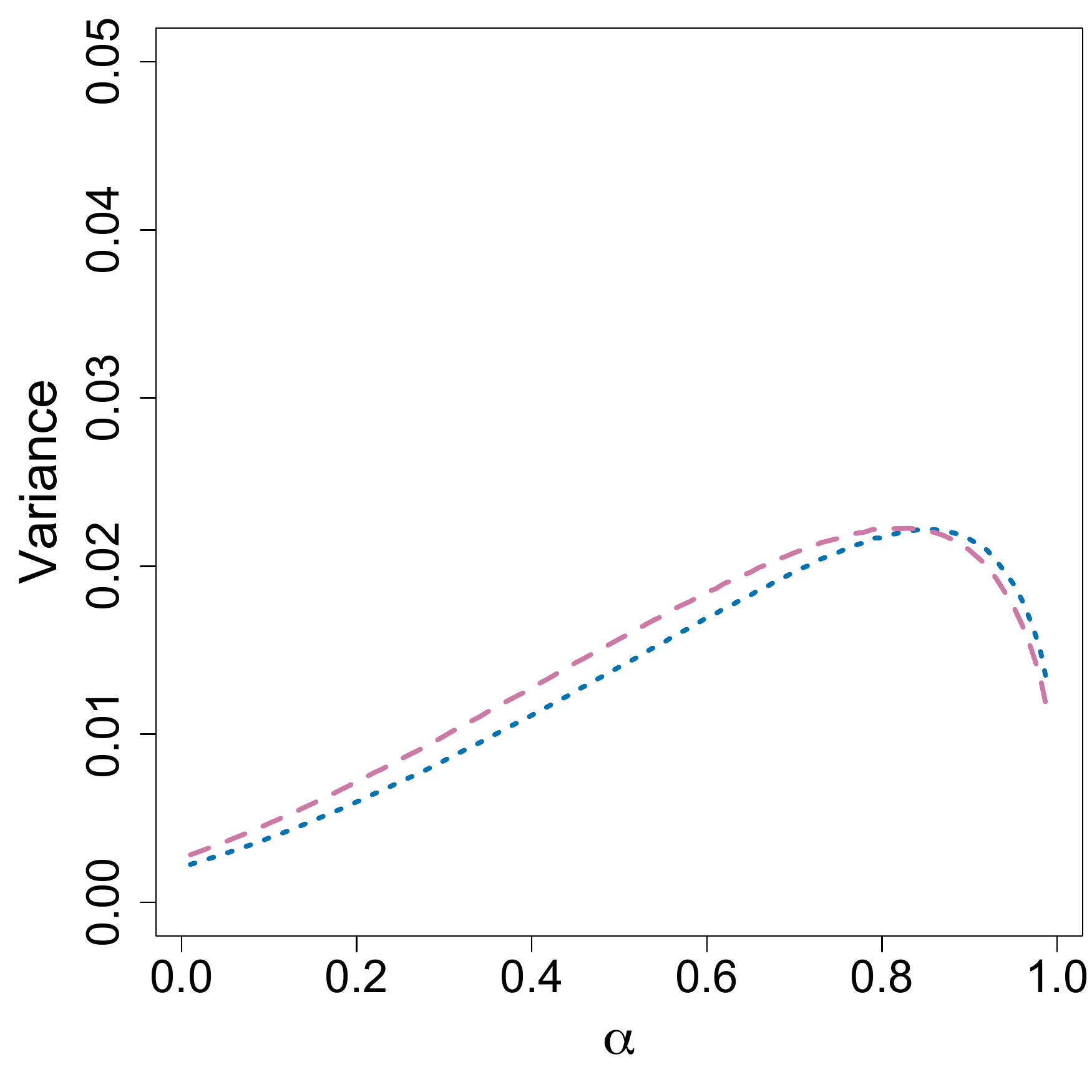} & \includegraphics[scale=.23]{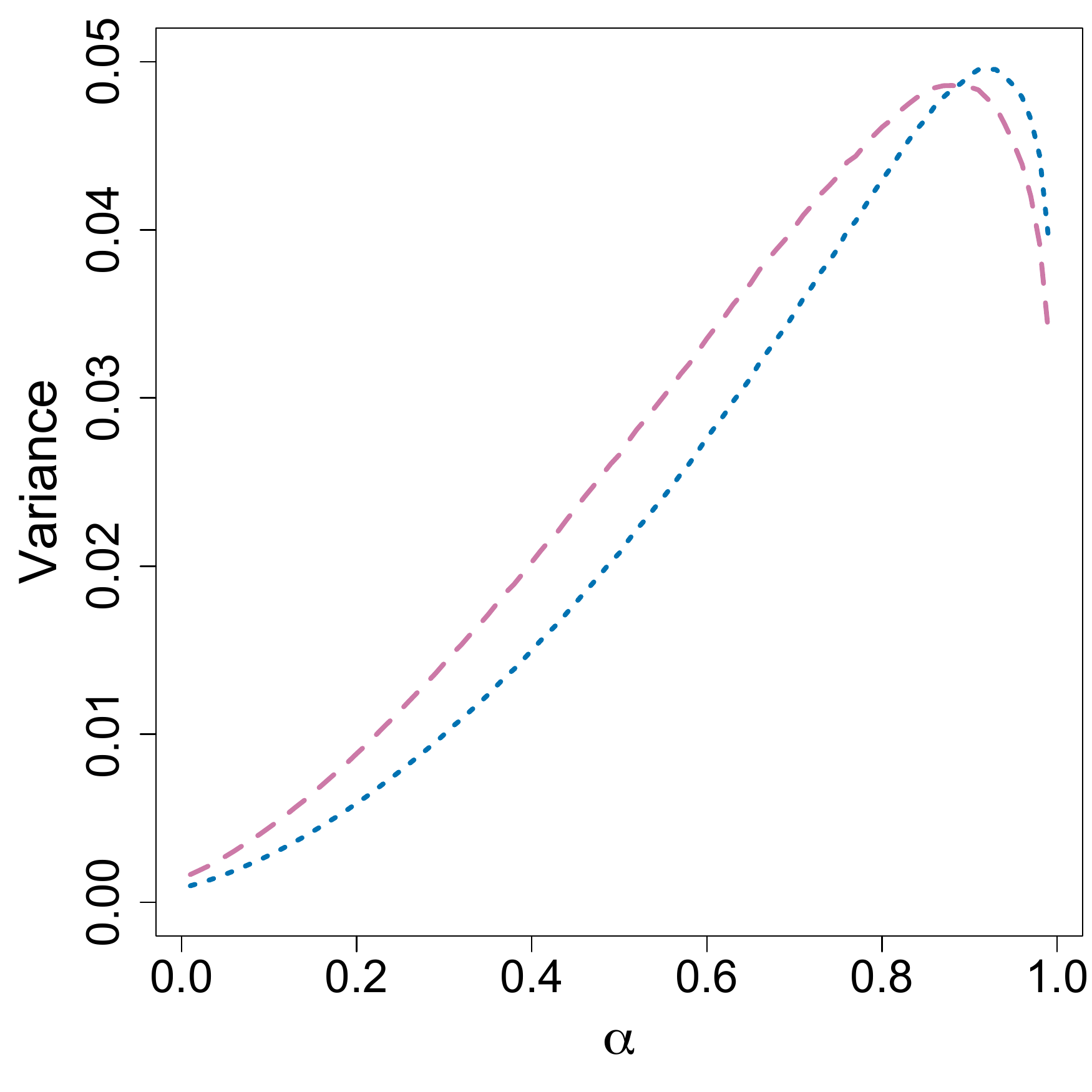} \\
& Categorical &\\
&&\\
\includegraphics[scale=.23]{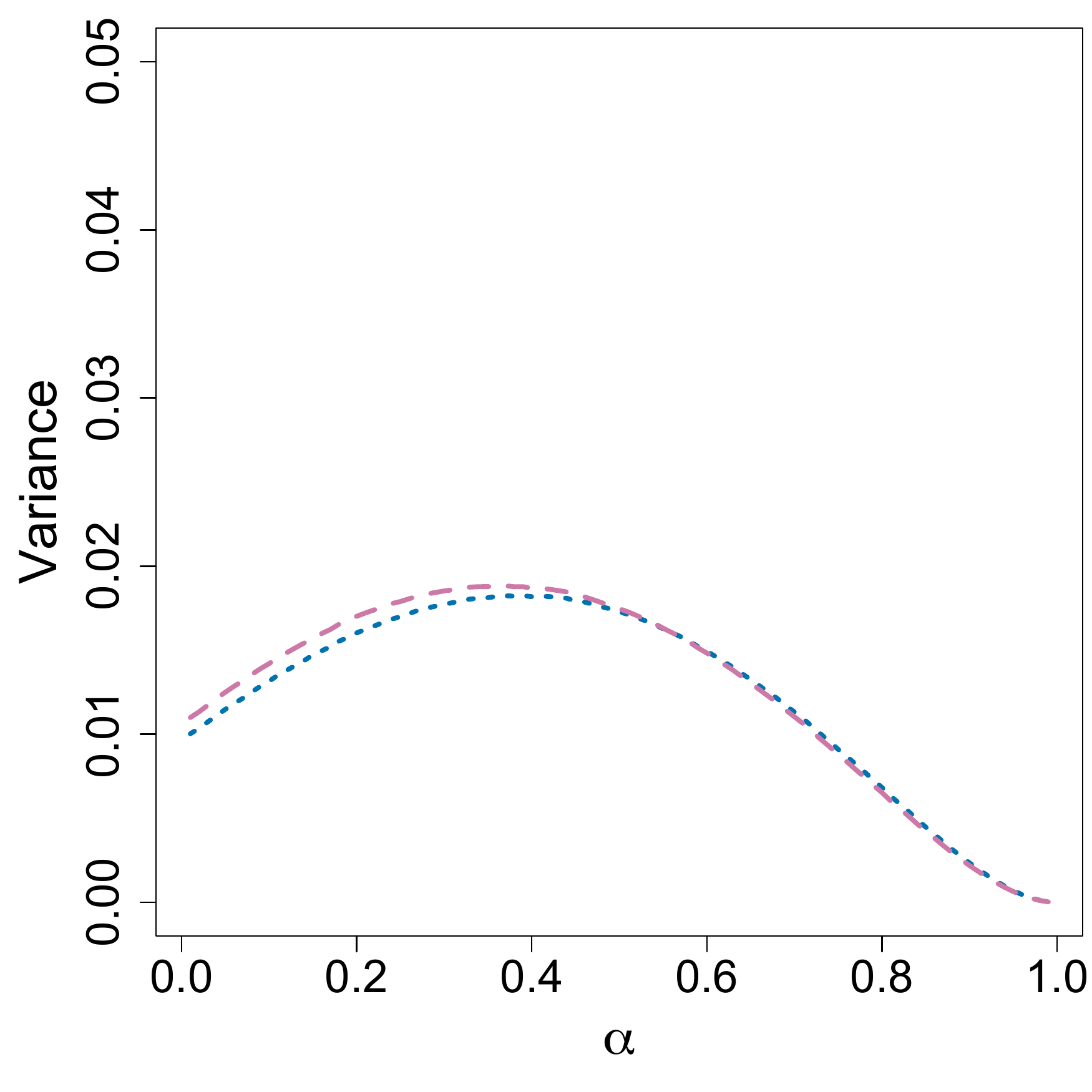} & \includegraphics[scale=.23]{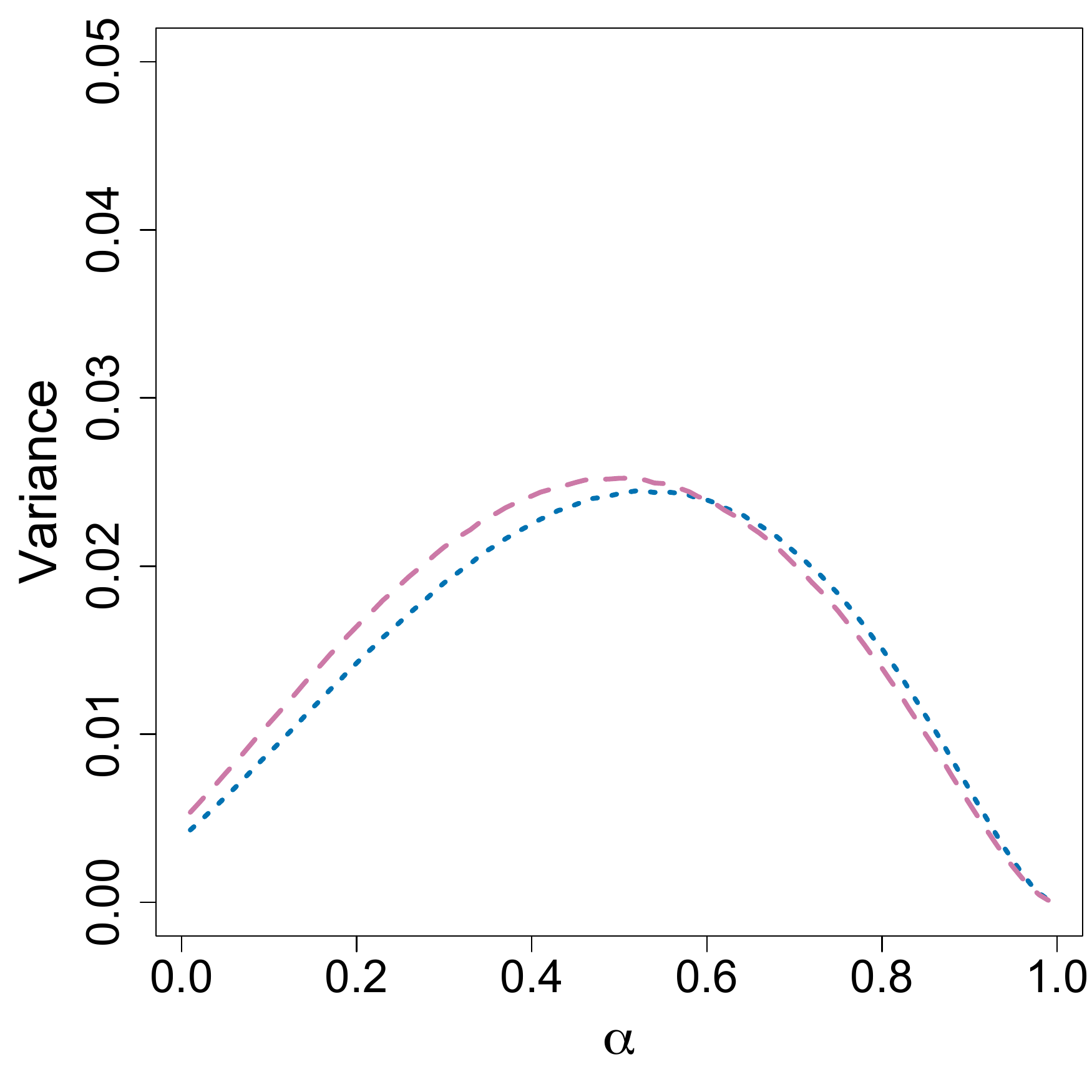} & \includegraphics[scale=.23]{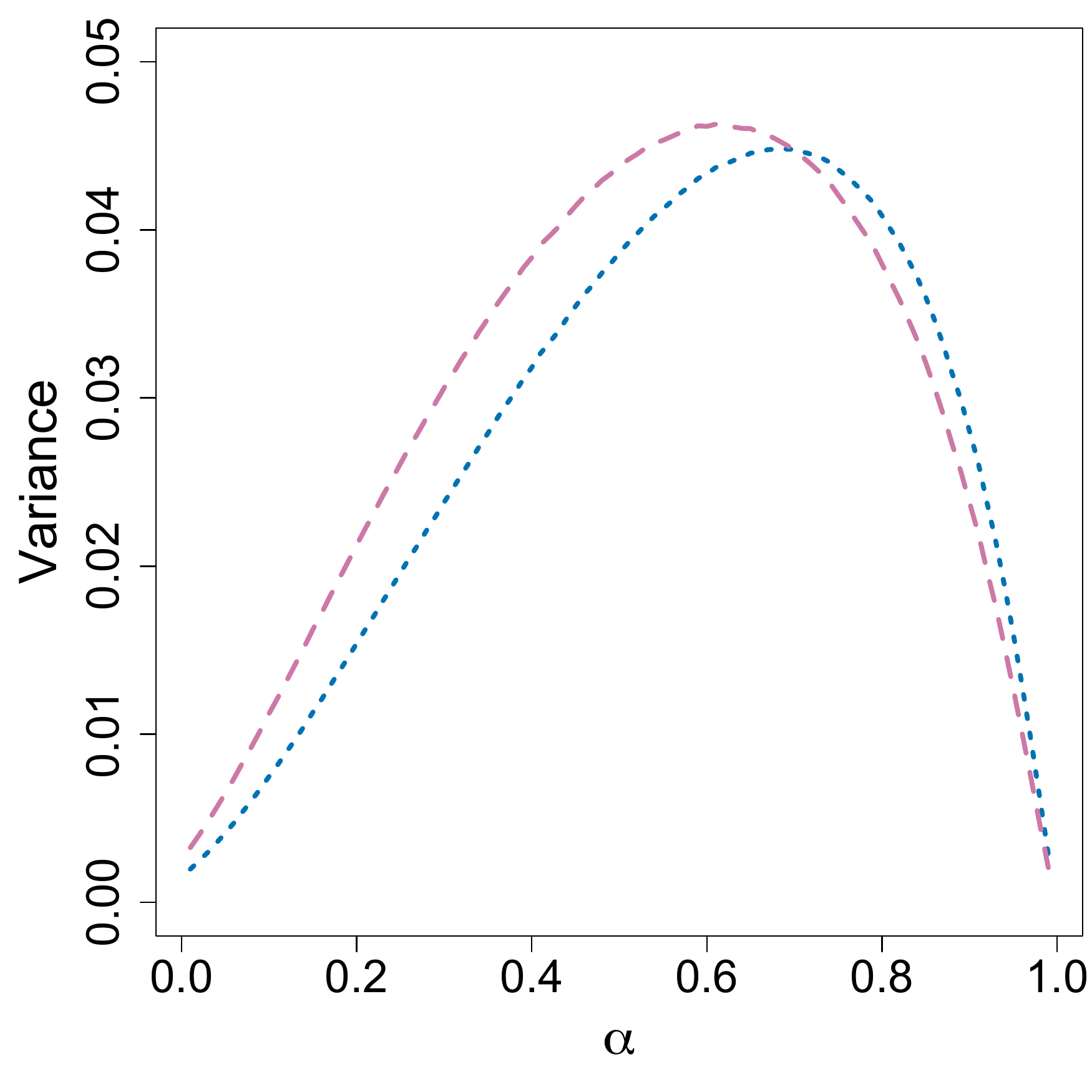} \\
& Gaussian &\\
   \end{tabular}
   \end{center}
   \caption{Estimation and variance of the customary and analytical estimators, for categorical outcomes.}
   \label{cat}
\end{figure}

I chose $p=3$ and $\bpi=(0.5, 0.2,0.3)'$ (a challenging scenario), and once again used a fine grid of $\alpha$ values. Results are shown in Figure~\ref{cat}, where results for the Gaussian scenario are included for comparison. We see that the relative behaviors of the two estimators are broadly similar for nominal data and Gaussian data, which suggests that my proposed methodology may substantially improve upon the customary methodology for all kinds of data.

\section{Application to experimentally observed data}
\label{realdata}

\subsection{Re-analysis of Krippendorff's categorical data}

Re-analysis of Krippendorff's nominal data using my proposed methodology yielded $\tilde{\alpha}=0.756$ and $\alpha\in(0.228,0.951)$. Leaving out row 6 gives $\tilde{\alpha}=0.866$ and $\alpha\in(0.370,0.981)$. All results for this dataset are collected in Table~\ref{kripp}.

Each width ratio is the ratio of the length of the jackknife interval to the length of the customary bootstrap interval. We see that the jackknife intervals are much wider than the bootstrap intervals, as expected.

The execution times in Table~\ref{kripp} (as well as in Table~\ref{liver} and Table~\ref{pm}) were obtained on a 3.6 GHz 10-core Intel Core i9 CPU, with both interval estimation methods parallelized over eight cores, and a bootstrap sample size of 2,000 for the customary procedure.

\begin{table}[ht]
   \centering
   \begin{tabular}{rccr}
   Estimate & 95\% Confidence Interval & Width Ratio & Execution Time\\\hline
  $\hat{\alpha}=0.743$ & $\alpha\in(0.459,1.000)$ & \multirow{2}{*}{1.34} & $<1$ s\\
  $ \tilde{\alpha} =0.756$ & $\alpha\in(0.228,0.951)$ & & 2 s\\\hline
  $\hat{\alpha}_{-6}=0.857$ & $\alpha\in(0.679,1.000)$ &  \multirow{2}{*}{1.90} & $<1$ s\\
  $ \tilde{\alpha}_{-6} =0.866$ & $\alpha\in(0.370,0.981)$ & & 2 s
         \end{tabular}
   \caption{Results from applying the customary and improved methodologies to Krippendorff's nominal data. }
   \label{kripp}
\end{table}

\subsection{Analyses of CDH data}

The data for this example, some of which are shown in Figure~\ref{ordinal}, are liver-herniation scores (in $\{1,\dots,5\}$) assigned by two coders (radiologists) to magnetic resonance images (MRI) of the liver in a study pertaining to congenital diaphragmatic hernia (CDH) \citep{cdh2020}, in which a hole in the diaphragm permits abdominal organs to enter the chest. The five grades are described in Table~\ref{grades}.

\begin{figure}[h]
   \centering
   \begin{tabular}{cccccccccccc}
   & $u_1$ &  $u_2$ & $u_3$ & $u_4$ & $u_5$ & $\dots$ & $u_{43}$ & $u_{44}$ & $u_{45}$ & $u_{46}$ & $u_{47}$\vspace{2ex}\\
   $c_{11}$ & 2 & 4 & 4 & 4 & 4 & $\dots$ & 2 & 1 & 2 & 1 & 1\\
   $c_{12}$ & 2 & 4 & 5 & 4 & 4 & $\dots$ & 2 & 1 & 2 & 1 & 1\\
   $c_{21}$ & 3 & 5 & 5 & 5 & 4 & $\dots$ & 2 & 2 & 2 & 1 & 1\\
   $c_{22}$ & 3 & 5 & 5 & 4 & 4 & $\dots$ & 2 & 2 & 2 & 1 & 1
   \end{tabular}
   \caption{Ordinal scores for MR images of the liver. Each coder scored each unit twice.}
   \label{ordinal}
\end{figure}

\begin{table}[h]
   \centering
   \begin{tabular}{cl}
   Grade & Description\\\hline
   1  & No herniation of liver into the fetal chest\\
2 & Less than half of the ipsilateral thorax is occupied by the fetal liver\\
3 & Greater than half of the thorax is occupied by the fetal liver\\
4 & The liver dome reaches the thoracic apex\\
5 & \parbox[t]{10cm}{The liver dome not only reaches the thoracic apex but also extends\\ across the thoracic midline}
   \end{tabular}
   \caption{Liver herniation grades for the CDH study.}
   \label{grades}
\end{table}

Each coder scored each of the 47 images twice, and so we are interested in assessing both intra-coder and inter-coder agreement. The results are shown in Table~\ref{liver}. We see that both intra-coder and inter-coder agreement are very nearly perfect. The jackknife interval widths are once again substantially wider than the bootstrap intervals because the jackknife approach properly accounts for uncertainty. And the running times are comparable for the two methods because the dataset is on the small side.

\begin{table}[ht]
   \centering
   \begin{tabular}{ccccc}
    & Estimate & 95\% Confidence Interval & Width Ratio & Execution Time\\\hline
   \multirow{2}{*}{Radiologist 1} & $\hat{\alpha}=0.979$ & $\alpha\in(0.950,1.000)$ & \multirow{2}{*}{1.42} & 2 s\\
   & $\tilde{\alpha}=0.979$ & $\alpha\in(0.923,0.994)$ & & 2 s\\\hline
   \multirow{2}{*}{Radiologist 2} & $\hat{\alpha}=0.987$ & $\alpha\in(0.968,1.000)$ & \multirow{2}{*}{2.56} & 2 s\\
   & $\tilde{\alpha}=0.987$ & $\alpha\in(0.916,0.998)$ & & 2 s\\\hline
   \multirow{2}{*}{Both}& $\hat{\alpha}=0.965$ & $\alpha\in(0.944,0.984)$ & \multirow{2}{*}{1.25}& 2 s\\
    & $\tilde{\alpha}=0.966$ & $\alpha\in(0.933,0.983)$ & & 2 s
         \end{tabular}
   \caption{Results from applying the customary and improved methodologies to the liver data. }
   \label{liver}
\end{table}

\subsection{Analyses of PM$_{2.5}$ data}
\label{almaty}

\begin{figure}[ht]
   \begin{center}
  \includegraphics[scale=.4]{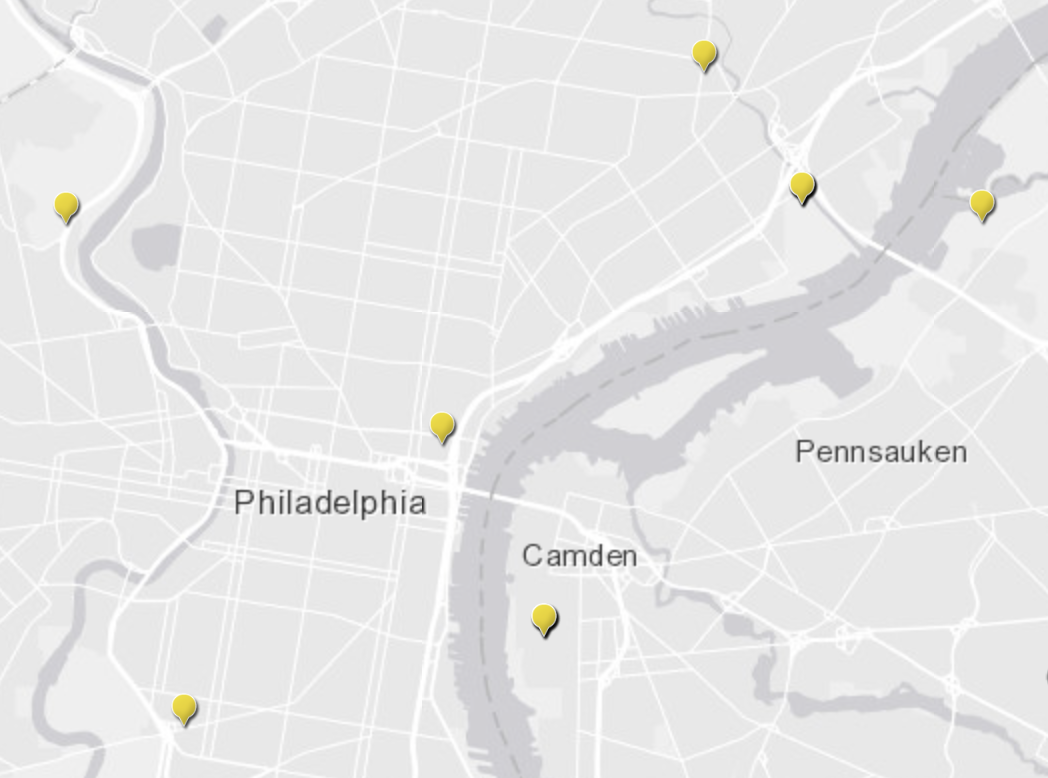}
     \end{center}
   \caption{The locations of the seven air quality monitors that generated the PM$_{2.5}$ data. I used the EPA's interactive map to create this image.}
   \label{philly}
\end{figure}

\begin{figure}[ht]
   \begin{center}
  \includegraphics[scale=.4]{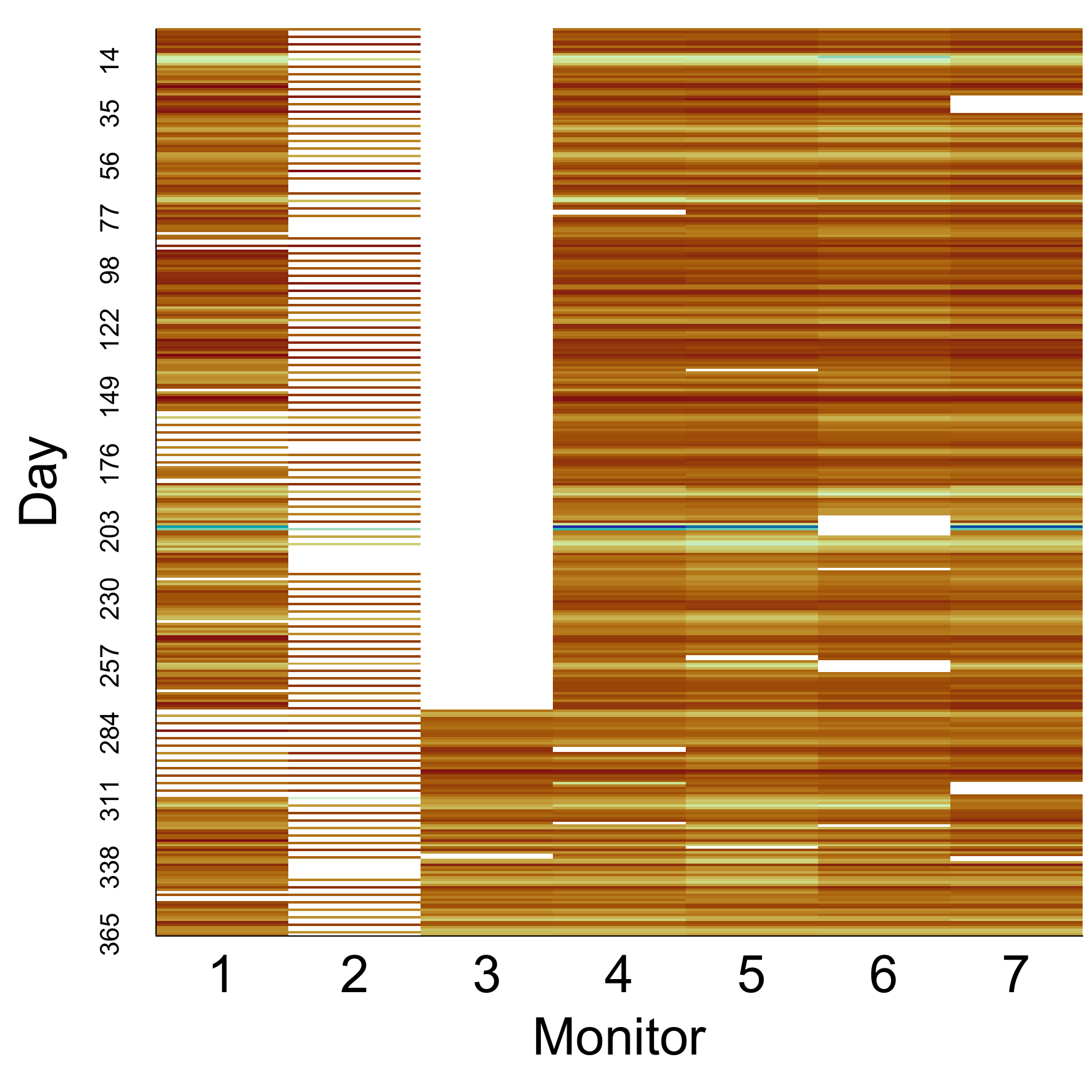}
     \end{center}
   \caption{The 2021 PM$_{2.5}$ data for Philadelphia, PA.}
   \label{pmdata}
\end{figure}

I created the PM$_{2.5}$ dataset from data collected during 2021 by the EPA (\url{https://tinyurl.com/2p93czeh}) via seven monitors in or near Philadelphia, Pennsylvania (see Figure~\ref{philly}). Specifically, the data matrix comprises exactly 365 rows, one row for each day of 2021. The row corresponding to a given day comprises that day's mean PM$_{2.5}$ concentrations (in micrograms per cubic meter) for the seven monitors. An image plot of the dataset is shown in Figure~\ref{pmdata}. We see extensive missingness for monitors 2 and 3, yet the dataset is large (1,937 measurements) and dense (76\%). The horizontal streaking in the image plot suggests that an analysis of these data will reveal considerable spatial agreement among the seven monitors.

My analysis yielded the results shown in Table~\ref{pm}. Since this dataset is quite large, it is not surprising that the customary method and my proposed method produced nearly identical results. (Although the jackknife interval is 31\% wider than the bootstrap interval, both intervals are narrow.) Nor is it surprising that the execution times were six seconds and nearly two minutes, respectively: for a dataset so large, the cubic time complexity of the jackknife procedure begins to show itself conspicuously. In this respect it is fortunate that the customary methodology yields high-quality inference for large datasets.

\begin{table}[ht]
   \centering
   \begin{tabular}{ccccc}
    Estimate & 95\% Confidence Interval & Width Ratio & Execution Time\\\hline
   $\hat{\alpha}=0.859$ & $\alpha\in(0.841,0.875)$ & \multirow{2}{*}{1.31} & 6 s\\
   $\tilde{\alpha}=0.859$ & $\alpha\in(0.835, 0.880)$ & & 1 min 53 s
         \end{tabular}
   \caption{Results from applying the customary and improved methodologies to the Philadelphia PM$_{2.5}$ data. }
   \label{pm}
\end{table}

These results suggest that a single air quality monitor may suffice for the study region, which could save taxpayer dollars that are currently being spent to purchase, install, and maintain the monitors. Perhaps further analyses of this sort for other cities would lead to similar conclusions and possibly more savings. And one can imagine how these and similar data could be used to measure agreement patterns not only for a whole year but also within a given year (by season, for example) or across years.

\section{Discussion}

In this article I revealed inferential challenges faced by the customary methodology for Krippendorff's $\alpha$ agreement measure. Specifically, the customary point estimator is biased downward, and the bias increases in magnitude rather substantially as the number of coders increases. The customary procedure for interval estimation is also wanting, typically yielding unacceptably low coverage rates for smaller samples.

I then investigated a number of candidate point estimators, and ultimately chose the so called analytical estimator owing to that estimator's bias and mean squared error profiles. I employed the analytical estimator to develop a jackknife approach to interval estimation. Although potentially burdensome computationally, the jackknife procedure produces confidence intervals having very nearly the nominal coverage rates.

In the final section of the paper I compared and contrasted the customary methodology and my proposed methodology as applied to three experimentally observed datasets. The first dataset was previously analyzed by Krippendorff. The latter two datasets---which pertain to an imaging study in congenital diaphragmatic hernia and the spatial pattern of PM$_{2.5}$ concentrations in and near Philadelphia, Pennsylvania---represent novel applications of Krippendorff's $\alpha$.

\section*{Author contributions}

John Hughes conceived the project, performed all simulations and data analyses, created the PM$_{2.5}$ dataset, and wrote the manuscript.

\section*{Acknowledgments}

I thank Hyunok Choi for helpful discussions regarding air quality monitoring.

\bibliography{refs}
\bibliographystyle{apalike}

\end{document}